\documentclass[11pt]{article}\pdfoutput=1
\usepackage{amsmath,amsfonts,amssymb,amscd}
\usepackage{pstricks,pst-node}
\usepackage[small,bf,hang]{caption}
\usepackage{tensor}
\usepackage{slashed}
\usepackage{amsmath}
\usepackage{latexsym,epsfig}
\usepackage{arydshln}
\usepackage{extarrows}
\usepackage[utf8]{inputenc}
\usepackage[linktoc=all,hidelinks]{hyperref}
\usepackage{cite}

\usepackage{graphicx} 
\usepackage{wrapfig}
\usepackage{lipsum}
\usepackage{caption}   
\graphicspath{{Images/}}

\usepackage{tikz}
 \usetikzlibrary{arrows.meta,calc}

\usepackage{amsthm,tikz-cd,accents}

\usepackage[makeroom]{cancel}

\usepackage{color}

\def\hybrid{
        \topmargin -20pt
        \oddsidemargin 0pt
        \headheight 0pt \headsep 0pt
        \textwidth 6.25in 
        \textheight 9.5in 
        \marginparwidth .875in
        \parskip 5pt plus 1pt \jot = 1.5ex}

\hybrid

 \csname
@addtoreset\endcsname{equation}{section}

\def\moth{\mathsurround=0pt}
\newdimen\zo \zo=0pt

\def\tick{\leaders\hrule height 0.5ex depth 0pt \hskip 0.5pt}
\def\upboxfill{$\moth \setbox\zo\hbox{\tick}%
  \hskip 3pt\hbox to 0pt{$\tick$\hss}\hrulefill \hbox to 7.5pt{$\tick$\hss}$}

\def\dtick{\leaders\hrule height .34pt depth 0.5ex \hskip 0.5pt}
\def\downboxfill{$\moth \setbox\zo\hbox{\dtick}%
  \hskip 2pt\hbox to 0pt{$\dtick$\hss}\hrulefill \hbox to 2pt{$\dtick$\hss}$}

\def\bec{\begin{center}}
\def\ec{\end{center}}

\def\be{\begin{equation}}
\def\ee{\end{equation}}
\def\bea{\begin{eqnarray}}
\def\eea{\end{eqnarray}}
\def\ba{\begin{array}}
\def\ea{\end{array}}

\begin{document}

\begin{titlepage}
\rightline{}
\rightline{May 2026}
\begin{center}
\vskip 2cm
{\Large \bf{Lecture Notes on  \\[1ex] 
Statistical Physics and Neural Networks  }
}\\
\vskip 1.2cm


  \vskip 1.5cm
 {\large {Olaf Hohm  }}
\vskip 1cm

{\it  Institute for Physics, Humboldt University Berlin,\\
 Zum Gro\ss en Windkanal 2, D-12489 Berlin, Germany}\\
\vskip .1cm

ohohm@physik.hu-berlin.de

\vskip 2cm
{\bf Abstract}

\end{center}


\noindent
\begin{narrower}

 These lecture notes introduce some topics of classical statistical physics, 
particularly those that are relevant for neural networks and  deep learning. 
Statistical physics is treated  as a branch of probability theory or statistics, 
with the goal of making concepts such as phase transitions and the renormalization group accessible 
to readers without prior knowledge of physics. 
We introduce  the Boltzmann-Gibbs distribution and the thermodynamic potentials 
on a finite configuration space, notably for  Ising spins and spin-glass models on a lattice, 
and then define  phase transitions as discontinuities that arise in the limit that 
the number of lattice points goes to infinity. We further introduce  Hopfield networks and Boltzmann machines, which are governed 
 by the same energy function as  spin-glass models, 
 and discuss the learning algorithm for restricted Boltzmann machines. In this algorithm  hidden neurons are 
 integrated out as in the renormalization group. 
 Finally,  modern deep learning is introduced, whose early developments were in part 
 motivated by restricted Boltzmann machines 
 in that they carry many layers of hidden neurons. 
 A description of  large language models is given.

\end{narrower}

\end{titlepage}

\tableofcontents


\newpage

\section{Introduction}

These lecture notes are based on a master level course on statistical physics at Humboldt University Berlin. More precisely, 
they cover about 30--40\%  of that course, focusing on the classical statistical mechanics of phase transitions and the method of the 
renormalization group (RG)  together with an introduction to neural networks and the basics underlying the modern artificial intelligence 
(AI) revolution such as large language models.

My motivation to present and write up  this material was  three-fold: First, the relevance 
of statistical physics concepts for neural networks is evident already in its vocabulary, including such terms as 
temperature,  entropy and Boltzmann machines. In fact, the Physics Nobel Prize 2024 was awarded to 
John Hopfield and Geoffrey Hinton for discoveries  
``that enable machine learning with artificial neural networks" \cite{Nobel2024}, and I felt that 
physics students may reasonably want to learn about this in a statistical physics course. 
Second, there is a growing literature proposing concepts  from quantum field theory (QFT) and the closely related 
statistical physics (and statistical field theory)  for neural networks, notably phase transitions and 
RG techniques, see, e.g., \cite{Mehta:2014xqf,Tishby,Koch,Lietal,Iso:2018yqu,Koch:2019fxy,Halverson:2020trp,Erdmenger:2021sot,Cotler:2022fze,Berman:2022uov,
DeepLearningbook,Berman:2023rqb,Berman:2024pax,Sun:2025dey}. 
QFT is famously opaque, but whichever of its techniques may turn out to be  useful for neural networks 
it plausibly shares with statistical physics. Since  the latter  is much easier to learn  I decided 
 to try to  produce self-contained lecture notes on these topics 
that do not assume much  prior  knowledge of physics. 
Third, and most importantly, I wanted to  learn  myself about the technical details of neural networks and AI.

The structure of these lecture notes is as follows. 
Section 2 gives an introduction to the basics of classical statistical physics \cite{Schwabl,PhysicsandInfo}.  
One is given a configuration 
space of states or events together with a  real-valued function (observable) on it. This function, which  we will call \textit{energy}  
without worrying about the  physical meaning,  defines a probability distribution, the Boltzmann-Gibbs distribution 
depending on a parameter we will call \textit{temperature}, again without worrying about the physical meaning. 
The concrete examples we study are the  Ising model and its generalizations called spin-glass models. 
These are defined on a finite set of $N$ points, which we can think of as being arranged into a cubic lattice. 
A point in configuration space is given by an assignment of the numbers $+1$ or $-1$ to each lattice point. 
In the  original Ising model these configurations were meant to  describe 
magnetic moments with spin up or spin down, but they  may also describe the activations of neurons 
or the paths  of a one-dimensional random walk ($-1$ a step to the left, $+1$ a step to the right). 
Since  the configuration spaces are finite 
we will only need simple (albeit multivariable) calculus together with linear algebra and a basic understanding 
of probability theory (for which I provide a quick introduction in the appendix). 
Finally, phase transitions will be studied as singularities that appear in the thermodynamic limit in which the 
number of lattice points goes  to infinity: $N\rightarrow \infty$.

 In section 3 the renormalization group (RG) will be introduced as a tool to identify 
 phase transitions. Historically, RG techniques  first arose in QFT and high-energy physics \cite{Gell-Mann:1954yli,Wilson:1969zs}, 
 but they are much easier to understand in the statistical physics of, say,  
 the Ising model 
 \cite{Kadanoff:1966wm,Wilson1,Wilson2,Goldenfeld}. 
 One simply asks for the probability distribution for a subset of configurations (say assignments of $\pm 1$ 
 to a subset of lattice points) \textit{irrespective} of what the remaining configurations are.  This so-called marginalized 
 probability distribution is obtained by summing over the configurations we wish to ignore and is again a 
 Boltzmann-Gibbs distribution, but for a new energy function. One says that the configurations or degrees of freedom 
 that were summed over have been `integrated out' 
 (in statistical field theory and QFT these are actual  integrals, albeit in QFT ill-defined ones). 
Considering  a suitably general class of energy functions this integrating out of degrees of freedom, called  
 RG transformation,  can be viewed as 
 a transformation of the `coupling constants' (coefficients appearing in the energy function). 
Iterating the RG transformations one  has an `RG flow' in the (generically infinite-dimensional) space of coupling constants. 
These concepts will be illustrated for the Ising model in one and two dimensions. Although any subset of degrees of freedom 
 can be integrated out, one typically chooses them in such a way that the remaining degrees of freedom form again a 
 simple lattice but with an increased (rescaled)   lattice spacing. The RG flow thus successively coarse-grains, 
 so that eventually microscopic details become unimportant.  
 A critical RG fixed point signals a second-order 
 phase transition, which is `scale invariant' in that the so-called correlation length goes to infinity.

After this introduction to  textbook statistical physics, in sec.~4 we turn to neural networks 
and Boltzmann machines \cite{Hopfield,Hinton1,HintonNobel}. 
The Hopfield network is defined by a set of $N$ neurons, which each can be in the state 
$+1$ or $-1$. Moreover, there are  symmetric weight factors  between any two neurons.  This neural network is 
subject to a dynamical time evolution by which at each time step the state of a particular neuron is set to $+1$ if the weighted 
input from all other neurons is positive and otherwise set to $-1$. Under this evolution the configurations 
flow to a local minimum of an energy function. Intriguingly, this energy   is precisely the spin-glass energy of sec.~3,
just renaming  Ising spins  into  neurons and 
coupling constants into  weights.  
Boltzmann machines are then defined as such Hopfield networks 
where the time evolution becomes probabilistic, governed by a temperature parameter. After some time the Boltzmann machine 
reaches thermodynamic equilibrium, which means that the probability of finding it 
in a given neural configuration is given by the Boltzmann-Gibbs  distribution for the spin-glass model. 
We also study a learning algorithm which solves the inverse problem of finding  weights so that a 
given distribution, say obtained from a data set, is approximately described by 
the Boltzmann machine. 
Finally, we turn to restricted Boltzmann machines, for which this learning algorithm becomes much more efficient. 
One has hidden and visible neurons, and only the latter are `observable' in the sense of being  directly related  to 
the data distribution that the Boltzmann machine should  describe.  There is a natural interpretation 
of this in terms of integrating out the hidden neurons as in RG.

In sec.~5 we turn to modern \textit{deep learning}. While the  neural networks arising  here are only loosely 
connected to the Hopfield networks of (restricted) Boltzmann machines, the core idea of deep learning is rooted 
in the hidden neurons. One introduces many hidden layers of neurons  that do not participate directly in the inputs and outputs of 
a so-called  feedforward network but which presumably  encode hidden regularities and correlations. 
After introducing these neural networks 
we explain the learning algorithm called backpropagation to adjust its weights \cite{HintonBackProp}. 
Finally, we give a simplified description of the deep learning architecture, called the transformer, 
which  yields the large language models (LLMs) that are at the heart of the current AI revolution \cite{Transformer}. 
Even though 
in these developments the importance  of statistical physics is less obvious, there are various speculations on how 
statistical physics and QFT might illuminate features of LLMs such as the famous \textit{scaling laws}  that govern their rapid 
improvement \cite{Kaplan:2020trn}. These I will briefly discuss in the outlook sec.~6.

\section{Basics of Statistical Physics} 

We introduce the core probability to be used, the Boltzmann-Gibbs  distribution, and other statistical concepts with naming schemes from physics 
such as temperature and entropy. 
Then we define phase transitions and illustrate them with Ising-type spin models. 

\subsection{Boltzmann Distribution}

We consider the finite set ${\cal X}$, the \textit{configuration space} or \textit{space of events} or \textit{sample space},\footnote{As a rule, any respectable object  should have at least three names.} whose elements we call configurations 
and typically denote by $x\in {\cal X}$. 
Specifying a configuration  $x\in {\cal X}$ gives a complete microscopic description of the system. 
The number of elements of ${\cal X}$ we denote by $n=|{\cal X}|$. Often, $x$ will be given as a tuple 
of $N$ variables, $x=(x_1,\ldots, x_N)$, each taking a finite number of real values.
Moreover, we consider \textit{observables} or \textit{random variables}. These are  real-valued functions 
 \be 
  {\cal O} :{\cal X}\rightarrow \mathbb{R} \;. 
 \ee
We assume that there is one distinguished observable, the \textit{energy function} $E(x)$.

The Boltzmann-Gibbs  probability distribution (or Boltzmann distribution for short) 
is defined in terms of the energy function $E(x)$ by 
 \be\label{BoltzmannP} 
  P_{\beta}(x) := \frac{1}{Z(\beta)} \,e^{-\beta E(x)} \;, 
 \ee
 where the normalization factor is also known as the \textit{partition function}: 
  \be\label{partitionfunction} 
   Z(\beta):= \sum_{x\in {\cal X}} e^{-\beta E(x)} \;, 
  \ee
and $\beta=\frac{1}{kT}$ is a real parameter, usually assumed to be positive, called the inverse temperature or \textit{coolness} 
(and $k$ is there for reasons of nostalgia). It should be emphasized that, as we will see, the partition function contains essentially 
all thermodynamic information.  
Eq.~(\ref{BoltzmannP}) indeed defines a probability distribution on ${\cal X}$: for any $\beta$, the function $P_{\beta}$ obeys 
 \be
  \forall x\in {\cal X}: \; P_{\beta}(x)\geq 0\;, \qquad \sum_{x\in {\cal X}} P_{\beta}(x)=1\;, 
 \ee
which then implies $\forall x\in {\cal X}: 0\leq P_{\beta}(x)\leq 1$.  

We can define the usual statistical quantities (see the appendix) like the expectation value of an observable:
 \be\label{expectationValueeee} 
  \langle {\cal O} \rangle_{\beta} := \sum_{x\in {\cal X}} {\cal O}(x)P_{\beta}(x)
  = \frac{1}{Z(\beta)} \sum_{x\in {\cal X}} {\cal O}(x)e^{-\beta E(x)}\;. 
 \ee
Important examples are the two-point and higher-point correlation 
functions defined for a configuration space of variables $x_i$, $i=1,\ldots, N$, 
taking a finite number of values as, e.g., $x_i\in\{+1,-1\}$. 
The two-point function is the expectation value  of the observable 
\be\label{2pointfunc} 
  {\cal O}_{ij}(x):= x_i x_j\;. 
 \ee 
 The expectation value  $\langle {\cal O}_{ij}\rangle$ one normally just writes as 
  \be\label{2poinsfuncr333} 
   \langle x_ix_j\rangle = \frac{1}{Z(\beta)} \sum_{x\in {\cal X}} x_i x_j \, e^{-\beta E(x)} \;, 
  \ee  
 although this notation 
is misleading, suggesting a dependence on $x$ that does not exist: the result, obtained by summing over all 
$x$ configurations, depends (apart from $\beta$ and possible coefficients or coupling constants entering the energy function) 
only on $i, j$, which we can think of as fixed but arbitrary. 
The two-point function contains important statistical information, as for instance the correlation length to be used below, 
c.f.~(\ref{TwoPointFunctionCorr}). Higher order correlation functions are defined analogously as the expectation values 
of observables defined as higher order monomials in the $x_i$.

\bigskip

  \noindent\textit{High-temperature limit:}\\[0.5ex] 
In order to understand the significance of the temperature parameter $\beta$ we consider the obvious limits: 
For high temperatures $T\rightarrow \infty$ we have $\beta\rightarrow 0$ and thus for any $x$
 \be
  \lim_{\beta\rightarrow 0} P_{\beta}(x) =  \lim_{\beta\rightarrow 0} \frac{1}{\sum_{x'\in {\cal X}} e^{-\beta E(x')} } \,e^{-\beta E(x)}
  =\frac{1}{\sum_{x'\in {\cal X}}1} = \frac{1}{|{\cal X}|} = \frac{1}{n} \;. 
 \ee
Thus, the Boltzmann distribution becomes the \textit{uniform probability distribution} that assigns any point in the 
configuration or sample  space the same probability, which is $\frac{1}{n}$ if there are $n$ configurations  or elementary events. 
Thus, at high temperature all configurations or elementary events are equally likely. 

\medskip
  \noindent\textit{Low-temperature limit:}\\[0.5ex] 
 The low temperature limit $T\rightarrow 0$ or $\beta\rightarrow \infty$ is more subtle. We need to define the \textit{space of  
 ground states}, i.e., the space of all configurations that minimize the energy function: 
  \be\label{Groundstates} 
   {\cal X}_0:=\Big\{ x'\in {\cal X}\,\Big| \,\forall x\in {\cal X}:\,E_0:=E(x') \leq E(x) \Big\}\;. 
  \ee
In words, ${\cal X}_0$ consists of all configurations in ${\cal X}$ that minimize the energy, with minimal 
energy $E_0$.  Since we are working with finite sets there must be at least one such $x$, but there could be several,  
and it could also be that all $x$ belong to ${\cal X}_0$ (in which case the  energy $E$ would be constant, 
and the Boltzmann distribution would be the uniform distribution). In order to take the limit we now define the shifted energy 
 \be\label{shiftedENERGY} 
  \Delta (x): = E(x)-E_0 \geq 0\;, 
 \ee
so that  
 \be\label{discriminator} 
  \Delta(x)=0 \quad \Leftrightarrow \quad x\in {\cal X}_0\;. 
 \ee
In the partition function (\ref{partitionfunction}) we can then split the sum as 
 \be
 \begin{split} 
   Z(\beta)&= \sum_{x\in {\cal X}} e^{-\beta E(x)} = e^{-\beta E_0} \sum_{x\in {\cal X}} e^{-\beta \Delta (x)} 
   = e^{-\beta E_0}\Big(  \sum_{x\in {\cal X}_0} 1 + \sum_{x \notin {\cal X}_0} e^{-\beta \Delta (x)} \Big) \\
   &= e^{-\beta E_0}\big(|{\cal X}_0| + R_{\beta}\big)\;, 
 \end{split} 
 \ee
where we defined $R_{\beta}:=\sum_{x \notin {\cal X}_0} e^{-\beta \Delta (x)} $. Importantly, by (\ref{shiftedENERGY}) and  (\ref{discriminator}) we have 
the well-defined limit 
 \be
  \lim_{\beta\rightarrow \infty} R_{\beta}=0\;. 
 \ee
We can now compute the limit of the probability: 
 \be
   \lim_{\beta\rightarrow \infty} P_{\beta}(x) =  \lim_{\beta\rightarrow \infty} \frac{e^{-\beta E(x)}}{e^{-\beta E_0}\big(|{\cal X}_0| + R_{\beta}\big)}
   =  \lim_{\beta\rightarrow \infty} \frac{e^{-\beta \Delta(x)}}{|{\cal X}_0| }
   = \begin{cases}
      \frac{1}{|{\cal X}_0|}  & \text{for $x\in {\cal X}_0$}\\
      0 & \text{for $x\notin {\cal X}_0$}
      \end{cases}      \;. 
 \ee
Thus, the Boltzmann distribution becomes  the uniform distribution on the space of ground states and zero otherwise. 
At low temperatures the low-energy configurations dominate. The energy of a system tends to its minimum at 
low temperatures in the sense that the low-energy states have the highest probability.

 \bigskip

  \noindent\textit{Marginalizing the Boltzmann  distribution:}\\[0.5ex] 
Sometimes we will ignore certain configurations  and work with an effective probability 
distribution for the remaining ones. 
Say the configuration space is the cartesian product ${\cal X}\times {\cal Y}$, 
whose  elements are  $(x,y)$, with $x\in {\cal X}$ and $y\in {\cal Y}$. 
We then ask: what is the probability for a certain $x$ configuration \textit{irrespective} of what the 
$y$ configurations are? The answer is obtained  by marginalizing the probability distribution: 
 \be\label{Peff} 
   P_{\rm eff}(x) := \sum_{y\in {\cal Y}} P(x,y) = \frac{1}{Z} \sum_{y\in {\cal Y}}e^{-\beta E(x,y)}\;.  
 \ee
Defining 
 \be\label{effectiveEnergyyyy} 
  e^{-\beta E'(x)} := \sum_{y} e^{-\beta E(x,y)} \qquad \Leftrightarrow \qquad 
  E'(x) := -\frac{1}{\beta} \ln \sum_{y} e^{-\beta E(x,y)}\;, 
 \ee
we can write the partition function (\ref{partitionfunction}) as 
 \be
  Z(\beta) = \sum_{x,y} e^{-\beta E(x,y)} = \sum_{x} e^{-\beta E'(x)}\;, 
 \ee
in terms of which the marginal probability distribution (\ref{Peff}) reads 
 \be\label{PEFFFFF} 
  P_{\rm eff}(x) = \frac{1}{Z} \, e^{-\beta E'(x)}\;. 
 \ee
This takes the Boltzmann form, with respect to a new energy function $E'(x)$ for the remaining 
configurations $x$. Note also that for an observable ${\cal O}$ on  ${\cal X}\times {\cal Y}$ 
that depends only on $x$, its expectation value with respect to the full 
probability distribution $P(x,y)$ 
equals 
 \be
   \langle {\cal O} \rangle  = \sum_{x} {\cal O}(x) P_{\rm eff}(x) \;. 
 \ee
Thus, the marginal probability distribution obtained after eliminating variables is the unique distribution that reproduces the expectation values of all observables depending only on the remaining variables.
These concepts will be crucial for the renormalization group to be discussed below, 
where one says that the $y$ have been `integrated out'.

\bigskip

\noindent\underline{Thermodynamic Potentials:}
Once we know the partition function explicitly we know the probability distribution explicitly and can in principle compute 
any desired statistical quantity. To this end it is convenient to define the following thermodynamic potentials. 
The \textit{free energy} is defined by 
 \be
  F(\beta) := -\frac{1}{\beta} \ln Z(\beta)\;, 
 \ee
and since the $-\frac{1}{\beta}$ is mostly historical it is also convenient to define 
 \be
  \Phi(\beta)  := \ln Z(\beta)\;, 
 \ee
sometimes called the \textit{free entropy}. It is then elementary to see with (\ref{expectationValueeee}) 
that the \textit{internal energy}, defined as 
 \be
  U(\beta) := \langle E\rangle_{\beta}\;, 
 \ee
can be computed from the partition function as 
 \be
  U(\beta) = -\frac{\partial \Phi }{\partial \beta} =\frac{\partial}{\partial\beta}(\beta F) \;. 
 \ee 
Furthermore, the \textit{entropy} of the Boltzmann distribution, 
 \be
  S(\beta):= -\sum_{x\in {\cal X}} P_{\beta}(x)\ln P_{\beta}(x)\;, 
 \ee
can  be computed as 
 \be
  S(\beta) = \beta^2\frac{\partial F}{\partial \beta} \;. 
 \ee
See appendix \ref{Shannon} for the information-theoretic meaning of entropy due to Shannon and also \cite{Feynman}. 

\subsection{Thermodynamic Limit and Phase Transitions }

Phase transitions denote discontinuities that appear at certain fixed  temperatures $\beta=\beta_c$ in thermodynamic quantities 
such as the free energy $F$. As long as we are working on finite configuration spaces ${\cal X}$ with finite $n=|{\cal X}|$ (or a finite 
number $N$ of dynamical variables each taking a finite number of values) and 
assume that the energy function $E(x)$ is analytic there can be no such discontinuities, because the partition function (\ref{partitionfunction}) 
and any quantities derived from it are then analytic too.  
Rather, phase transitions can only arise in the \textit{thermodynamic limit} $N\rightarrow \infty$ (or, thinking of the 
finite number of degrees of freedom being inside a box of volume $V$, the infinite volume limit $V\rightarrow \infty$). 
Similarly, expectation values of observables, such as the two-point function (\ref{2poinsfuncr333}), are manifestly 
analytic for finite $N$, but in the large $N$ limit may show discontinuities or singularities. Indeed, the  \textit{correlation length} 
defined in terms of the two-point function in this limit, c.f.~(\ref{TwoPointFunctionCorr}) below, diverges at a phase transition.

Denoting the free energy defined above for a system with finite $N$ by $F_{N}(\beta)$ one typically has to rescale 
in order to have a well-defined thermodynamic limit. This yields the free energy density 
 \be\label{freenergydensity} 
  f(\beta):= \lim_{N\rightarrow \infty} \frac{1}{N} F_{N}(\beta)\;. 
 \ee
There is no longer any reason why the function obtained by such a limit should not have any discontinuities, 
and these, should they arise,  define the phase transitions. 

In practice, these discontinuities appear at isolated points $\beta= \beta_c=\frac{1}{kT_c}$. One classifies 
phase transitions as follows:\\[1.5ex] 
\underline{First-order phase transition:} \\
A phase transition at $\beta=\beta_c$ is first-order if $f(\beta)$ is continuous everywhere but not differentiable at $\beta=\beta_c$, 
i.e., $\frac{\partial f}{\partial \beta}\big|_{\beta_c}$ does not exist. 
\\[1.5ex] 
\underline{Second-order phase transition:} \\
A phase transition at $\beta=\beta_c$ is second-order if $f(\beta)$ is continuous and differentiable everywhere but its second derivative  
$\frac{\partial^2 f}{\partial \beta^2}\big|_{\beta_c}$ does not exist. 

\medskip
\noindent Obviously one may define phase transitions of yet higher order, but these do not tend to arise in practice 
and will not be considered here.

The following clarification is in order: If we take a bucket of water and put it outside on a cold winter day it will  freeze to solid ice once 
its temperature falls below $0^{\circ}$C. This is interpreted as a phase transition even though the bucket contains only a finite 
number $N$ of water molecules (say of order $N\sim 10^{25}$). Therefore,  thermodynamic quantities like the free energy are 
still analytic, but close to the critical temperature they exhibit a very sharp jump that for all practical purposes is like a 
discontinuity.  Studying the strict mathematical discontinuities of the $N\rightarrow \infty$ limit  is thus a method 
to identify those points with sharp jumps that signify a physical phase transition.

\subsection{Ising Model}

We now study  the Ising model \cite{Schwabl,PhysicsandInfo}. 
It was introduced to model magnets, consisting of a large number 
of molecules or atoms with a magnetic moment, which due to quantum mechanics is discrete (say with spin up or down). 
Later we take the Ising model as a model of neurons. 

The Ising model is defined on a $d$-dimensional cubic lattice of length $L\in \mathbb{N}$:  
 \be
  \Lambda_d:= \{1,2,3,\ldots, L\}^d \subset \mathbb{Z}^d\;. 
 \ee
Thus, the lattice points are the points in $\mathbb{R}^d$ with integer coordinates, where the integers run 
between $1$ and $L$. The volume of the lattice is $V=L^d$. See figure \ref{2Dlattice} below for a two-dimensional 
lattice. 

To each lattice point one assigns the number $+1$ (for spin up) or $-1$ (for spin down).  
Thus, labelling the lattice points by $i=1,\ldots, N:=L^d$, the dynamical variables are $\sigma_i=\pm 1$, and 
the configuration space is\footnote{The ${\sigma}$ 
are the variables that were previously denoted $x$, but it is considered to be in poor taste  
to denote spin variables by any other letter than $\sigma$.} 
 \be
  {\cal X}=\big\{{ \sigma} :=(\sigma_1,\ldots, \sigma_N)\,,\;\; N=L^d\big\}\;, 
 \ee
where we allow ourselves the short-hand  $\sigma$ for the $N$-tuple $(\sigma_1,\ldots, \sigma_N)$.  
The number of possible configurations  (the number of elements of ${\cal X}$) is then 
  \be
  n:=|{\cal X}|= 2^N = 2^{L^d}\;. 
 \ee

Having defined the configuration space we need to define the energy function to complete the description: 
 \be
  E({\sigma}) = -J \sum_{(ij)} \sigma_i \sigma_j - B\sum_i\sigma_i\;, 
 \ee
where $J$ is the coupling constant, and the sum in the first term is over (unordered) pairs of nearest neighbors $i,j$. 
This means that if, say, $i=3$ and $j=5$ are connected by an edge of the cubic lattice 
one includes the term $-J\sigma_3\sigma_5$. 
Otherwise one doesn't. 
The constant parameter $B$ models an external magnetic field.

\bigskip

\noindent\textit{Exact solution for the $d=1$ Ising model:}\\[0.5ex] 
The one-dimensional Ising model can be solved exactly, by which we mean that the partition function can be summed 
into an explicit function that allows one  to take the thermodynamic limit. 
This will show that the $d=1$ Ising model exhibits  no phase transition. 

We set $B=0$ and write the energy function of the $d=1$ Ising model as follows: 
 \be\label{B=0Isung}
  E({\sigma}) = -J\sum_{i=1}^{N} \sigma_i\sigma_{i+1}\;, \quad {\rm where} \qquad \sigma_{N+1}=\sigma_1\;, 
 \ee
i.e., we assumed periodic boundary conditions (the spins can be visualized as being on a circle). 
The partition function then reads 
 \be\label{Isingd=1sum} 
 \begin{split}
  Z_N(\beta) &= \sum_{\{\sigma_i=\pm 1\}}e^{-\beta E({\sigma})} =   \sum_{\{\sigma_i=\pm 1\}}e^{K\sum_{i}\sigma_i\sigma_{i+1}}\\
  &=\sum_{\sigma_1=\pm 1} \cdots \sum_{\sigma_N=\pm 1} e^{K\sigma_1\sigma_{2}}e^{K\sigma_2\sigma_{3}}\cdots 
  e^{K\sigma_{N-1}\sigma_{N}} e^{K\sigma_N\sigma_{1}}\;, 
 \end{split} 
 \ee
where $K=\beta J$.  
The trick is now to define the \textit{transfer matrix} $T$, which is the  2-by-2 matrix with components 
 \be
  T_{\sigma, \sigma'} = e^{K\sigma\sigma'} \;. 
 \ee
A typical sum appearing in (\ref{Isingd=1sum}) can then be written as a matrix product: 
 \be
  \sum_{\sigma_2=\pm 1} e^{K\sigma_1\sigma_{2}}e^{K\sigma_2\sigma_{3}} = (T^2)_{\sigma_1\sigma_3} \;. 
 \ee 
Thus, performing all sums over $\sigma_2,\ldots, \sigma_N$ we have 
 \be
   Z_N(\beta) = \sum_{\sigma_1=\pm 1}(T^N)_{\sigma_1\sigma_1} = {\rm Tr}(T^N)\;. 
 \ee
The problem has been reduced to computing the trace of $T^N$. This is easiest done 
by diagonalizing $T$ to $T'={\rm diag}(\lambda_+, \lambda_-)$ and recalling  that the trace is invariant: 
 \be\label{exactPart} 
  Z_N(\beta) = {\rm Tr}((T')^N) = \lambda_+^N + \lambda_-^N\;. 
 \ee
The two eigenvalues of $T$ are quickly found to be 
 \be
  \lambda_{\pm} = e^{K}\pm e^{-K} \;, 
 \ee
which back in (\ref{exactPart}) gives the exact partition function.

 Having found the exact partition function we can inspect  the thermodynamic limit $N\rightarrow \infty$.  
We can write  for the partition function $Z_N=\lambda_+^N\Big(1+\big(\frac{\lambda_-}{\lambda_+}\big)^N\Big)$ and hence 
 \be
  \frac{1}{N} \ln Z_N = \ln \lambda_+ + \frac{1}{N} \ln \Big(1+\big(\frac{\lambda_-}{\lambda_+}\big)^N\Big)\;, 
 \ee
so that due to $\lambda_+ >\lambda_-$  in the limit $\lambda_+ $ dominates. 
We  obtain for the free energy density (\ref{freenergydensity}) 
  \be\label{freenergydensity3324} 
  f(\beta)= \lim_{N\rightarrow \infty} \frac{1}{N} F_{N}(\beta) = -\lim_{N\rightarrow \infty} \frac{1}{N \beta}
 \ln Z_N = -\frac{1}{\beta} \ln(\lambda_+) \;, 
 \ee
or 
 \be\label{exactfreeenergylimit} 
   f(\beta)= -\frac{1}{\beta} \ln \big( 2 \cosh( K)\big) = -\frac{1}{\beta} \ln\big( 2 \cosh(\beta J)\big) \;. 
 \ee 
Since $ \cosh(K)$ is analytic and larger or equal to $1$, the free energy density is also analytic. 
Therefore: 
\textit{The one-dimensional Ising model does not exhibit a phase transition.}

\subsection{Spin-glass and Other Models}

\noindent\textit{Curie--Weiss Model:}\\[0.5ex]  
 A model similar to the Ising model, with dynamical variables $\sigma_i=\pm 1$, 
is the Curie--Weiss model, and this model does exhibit a phase transition. 
The discussion given here follows \cite{PhysicsandInfo}. 
The energy function of the Curie--Weiss model is given by 
 \be\label{CWenergy} 
  E({\sigma}) = -\frac{1}{N} \sum_{ij} \sigma_i\sigma_j-B\sum_{i=1}^{N} \sigma_i\;, 
 \ee
where now the sum runs over \textit{all} $\frac{1}{2}N(N-1)$ unordered pairs. What makes this model special, albeit perhaps 
somewhat unrealistic as a model for real materials, 
is that the energy can be written entirely in terms of the observable 
 \be
  m({\sigma}) := \frac{1}{N}\sum_{i=1}^{N} \sigma_i\;, 
 \ee
whose expectation value $\langle m \rangle_{\beta}$ is the so-called  \textit{magnetization} per spin. 
The energy (\ref{CWenergy})  becomes 
 \be
   E({\sigma}) = \frac{1}{2} -\frac{1}{2}N   m({\sigma})^2 - N B  m({\sigma})\;. 
 \ee
The factor $\frac{1}{2}$ is due to the square including a double sum that yields each pairing twice. 
Moreover, the sum yields the term $\frac{1}{2}(\sigma_1^2+\sigma_2^2+\cdots \sigma_N^2)=\frac{1}{2}N$
that is cancelled by the first term.

For the computation of the partition function one would then like to replace the sum over all ${\sigma}$ by a
sum over the $m$, which takes the values 
 \be\label{m(k)} 
  m=-1 + \frac{2k}{N}\;, \quad 0\leq k \leq N\;. 
 \ee
Indeed, the configuration ${\sigma}=(+1,+1,\ldots, +1)$ corresponds to $m=1$, ${\sigma}=(-1,-1,\ldots, -1)$
to $m=-1$ and any configuration obtained from this  by flipping $k$ spins corresponds to (\ref{m(k)}). 
Thus, we can replace the sum over ${\sigma}$ by a sum over $k=0,\ldots,N$, but 
one has to take into account  that there are several  ${\sigma}$ configurations that lead 
to the same $m$. A given configuration is determined by the number $N_+$ of $\sigma_i=+1$ or, equivalently, 
by the number $N_-$ of  $\sigma_i=-1$, as 
 \be
  N_++N_-=N\;, \qquad \text{where} \qquad m = \frac{N_+-N_-}{N} \;. 
 \ee
Solving these two equations one obtains 
 \be\label{solveNpm} 
  N_+ = \frac{N}{2}(1+m)\;, \qquad N_- = \frac{N}{2}(1-m)\;. 
 \ee 
The number of configuration with the same  $m$ is  given by the binomial 
 \be
  {\cal N}_N(m)= \binom{N}{N_+} 
  = \frac{N!}{N_-! \,N_+!} \;. 
 \ee 
(In appendix \ref{RandomWalk} essentially the same math is used for a one-dimensional random walk.) 
Thus, we can write the exact partition function as 
 \be
  Z_N(\beta, B) = e^{-\frac{\beta}{2}} \sum_{k=0}^{N} {\cal N}_N(m(k)) \exp\Big(\frac{N\beta}{2} m(k)^2 + N\beta B m(k)\Big)\;, 
 \ee
where $m(k)$ is given by (\ref{m(k)}).

\begin{figure}
    \centering
    \includegraphics[width=0.45\textwidth]{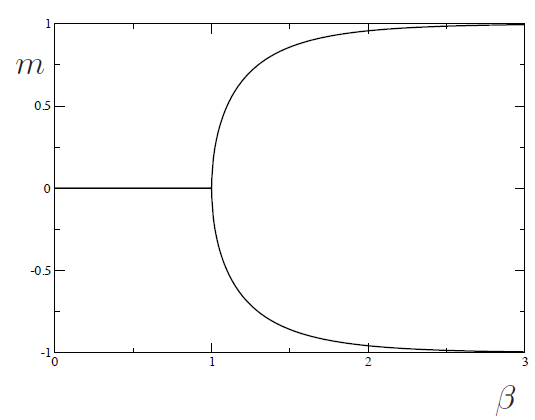}
    \caption{Phase transition of Curie-Weiss model (figure from  \cite{PhysicsandInfo})  }
    \label{CurieWeiss}
\end{figure}

The above  sum is an exact formula for the partition function, but we will not aim to compute it explicitly. 
Rather, we perform two approximations: First, we use Stirling's formula (\ref{Stirling}) 
together with  (\ref{solveNpm}), which after 
some straightforward algebra stablish the approximation 
 \be
  \frac{1}{N}\ln  {\cal N}_N(m) \ \simeq \ -\frac{1+m}{2} \ln \frac{1+m}{2}  -\frac{1-m}{2} \ln \frac{1-m}{2} \ =: \ s(m) \;, 
 \ee
and hence 
 \be\label{calNapprox} 
   {\cal N}_N(m) \ \simeq \  \exp\big(N s(m) \big)\;. 
 \ee 
Second, we approximate the sum by an integral, so that in total: 
 \be
  Z_N(\beta, B) \simeq e^{-\frac{\beta}{2}} \int_0^N dk \exp\Big(N\Big[s(m(k))+ \frac{\beta}{2} m(k)^2 + \beta B m(k)\Big]\Big)\;. 
 \ee
 We can now perform a substitution of integration variables: $k\rightarrow m(k)$, where using (\ref{m(k)}) we have $dm = \frac{2}{N}dk$, 
 and the integration range becomes $m\in [-1,1]$: 
 \be\label{approxpartition} 
  Z_N(\beta, B) \simeq e^{-\frac{\beta}{2}}\frac{N}{2} \int_{-1}^1 dm \exp\big(NG(m) \big) \;, 
 \ee
where  
  \be
    G(m):= s(m)+ \frac{\beta}{2} m^2 + \beta B m\;. 
  \ee
 Instead of trying to compute the integral exactly we will assume that it is dominated by the absolute maximum  of this function, 
 which obeys 
  \be\label{STEP135}  
   0=\frac{d G}{dm} = \frac{ds}{dm} + \beta m +\beta B = -\frac{1}{2}\ln \frac{1+m}{1-m} + \beta (m +B)\;. 
  \ee
 In order to simplify this equation we note the identity  $\frac{1+\tanh(x)}{1-\tanh(x)} = e^{2x}$, so writing 
  \be
   m = \tanh(\beta(m+B)) 
  \ee
 automatically solves (\ref{STEP135}). We will  solve this equation for $m(\beta)$ for  $B=0$. 
 Thus we seek the zeroes of the function 
  \be
   h(m):=  \tanh(\beta m) - m\;. 
  \ee
 This is still too hard to do exactly, but we can Taylor expand around $m=0$ to third order, 
  \be
   h(m) = (\beta -1)m -\frac{\beta^3}{3} m^3+\cdots\;. 
  \ee
 Thus, we want to solve 
  \be
    (\beta -1)m -\frac{\beta^3}{3} m^3  =0 \;. 
  \ee
 The solutions are 
  \be\label{twoms} 
   m=0\;, \qquad m=\pm\sqrt{\frac{3}{\beta^3}}\sqrt{\beta-1}\;,  
  \ee
 but the latter two are  only real for 
  \be
   \beta\geq 1\;. 
  \ee 
This indicates that there is a phase transition at $\beta=\beta_c=1$: For $\beta<\beta_c$ we have $m(\beta)=0$, 
but at the critical point $m(\beta)$ bifurcates into one of the two non-zero  solutions (\ref{twoms}), see figure \ref{CurieWeiss}. 
As $m(\beta)$ 
determines (under the assumed approximation) the partition function (\ref{approxpartition}) this 
discontinuity gives rise to a corresponding discontinuity of the partition function. Hence, there is a phase transition. 
More precisely,  the phase transition is second-order.

\bigskip

\noindent\textit{Spin-glass Models:}\\[0.5ex]   
 Spin-glass models were introduced to describe materials that are a mixture of two substances, with a large ratio between 
 them, say 1:100. For instance, the Edwards-Anderson model aims to describe such systems on the same configuration space 
 as the Ising model, but with an energy function given by 
  \be\label{spinglass} 
   E({\sigma}) = -\sum_{(ij)} J_{ij} \sigma_i\sigma_j - B\sum_{i}\sigma_i\;, 
  \ee  
 where in the first term one often  still assumes that only nearest-neighbor interactions are included, i.e., the coupling constants 
 $J_{ij}$ are only non-zero if the lattice points labelled by $i$ and $j$ are connected by an edge. The main novelty of this model 
 is the dependence on the coupling constants $J_{ij}$, so that the partition function will also depend on these: 
  \be
   Z_{N}(\beta, B, J_{ij}) = \sum_{{\sigma}}\exp\Big( \beta\sum_{(ij)} J_{ij} \sigma_i\sigma_j +\beta B\sum_{i}\sigma_i\Big)\;. 
  \ee
One often assumes that the $J_{ij}$ are drawn from a random distribution. 
 What makes spin-glass models much more subtle is that these constants could have both signs, which we denote by 
  \be
  \begin{split} 
   &J_{ij} > 0\;: \qquad {\rm ferromagnetic}\;, \\
  &J_{ij} < 0\;: \qquad {\rm anti-ferromagnetic} \;.
  \end{split} 
  \ee  
 This implies that the space of ground states (configurations that minimize the energy, c.f.~(\ref{Groundstates})) can be very rich, 
due  to the notion of \textit{frustration}. 

 \begin{figure}
    \centering
    \includegraphics[width=0.65\textwidth]{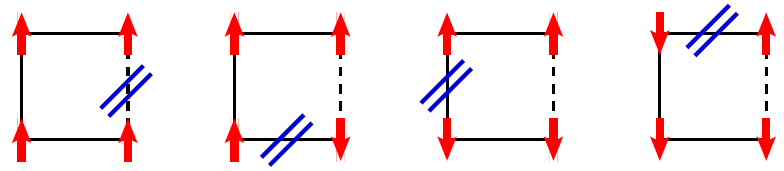}
    \caption{Frustration: For a  solid line we have $J=+1$, for a dashed line $J=-1$, and a crossed out line indicates 
    inconsistency  with minimizing the energy. (Figure from  \cite{PhysicsandInfo})  }
    \label{frustration}
\end{figure}

 To explain frustration consider $J_{ij}=\pm 1$ and $B=0$. The energy associated to an edge connecting $i$ and $j$  is then 
  \be
   \begin{split}
    J&_{ij}=1\,: \qquad E=-\sigma_i\sigma_j \;\quad \Rightarrow \quad \text{minimum for $\sigma_i=\sigma_j$}\\
    J&_{ij}=-1\,: \quad\; E=\sigma_i\sigma_j \qquad \Rightarrow \quad \text{minimum for $\sigma_i=-\sigma_j$} \;. 
   \end{split} 
  \ee
Put differently, in order to minimize the energy,  for  $J=+1$ the spins want to be parallel, while for $J=-1$ 
they want to be anti-parallel. If one considers a plaquette consisting of four lattice points, see figure \ref{frustration}, 
the minimal energy 
is $E=-4$ if all $J=\pm 1$ have the same sign, with the spins being all parallel (for $J=1)$ or alternating (for $J=-1$). 
However, if some $J$ are $+1$ and some $-1$ the minimal energy is the larger $E=-2$, with a degeneracy in configurations, 
provided 
 \be
  J_{12} J_{34} J_{13}J_{24} = -1\;.   
 \ee
This inability to assign spins to lower the energy down to the minimum of, say, the Ising model is aptly called 
\textit{frustration}.

\bigskip

\noindent\textit{High-temperature expansion:}\\[0.5ex]   
While the spin-glass models are very hard to analyze at low temperatures, where the intricate space of ground states dominates, 
at very high temperature one can apply perturbation theory, perturbing around $\beta=0$ or $T=\infty$. 
Considering  the energy function 
 \be
  E(\sigma) = -\frac{1}{2} \sum_{i,j=1}^{N} J_{ij} \sigma_i \sigma_j\;, \qquad J_{ij}=J_{ji}\;, \quad J_{ii}=0\;, 
 \ee
 we need to compute the partition function 
  \be
   Z(\beta, J) = \sum_{\sigma } e^{-\beta E(\sigma)} 
   = \sum_{\sigma_1=\pm 1} \cdots \sum_{\sigma_N=\pm 1} \exp\Big(\tfrac{\beta}{2}  \sum_{i,j} J_{ij} \sigma_i \sigma_j\Big)\;. 
  \ee
We illustrate the method of perturbation theory by Taylor expanding to second order in $\beta$: 
 \be\label{partitionfunctionspinglasssss} 
  Z= \sum_{\sigma } \Big(1 + \frac{\beta}{2}  \sum_{i,j} J_{ij} \sigma_i \sigma_j
  +\frac{\beta^2}{8} \sum_{i,j,k,l}J_{ij} J_{kl} \sigma_i\sigma_j \sigma_k \sigma_l +\cdots\Big)\;. 
 \ee
 
In order to evaluate this we need to perform the sums over all $\sigma $ configurations for various powers of $\sigma_i$, for instance: 
  \be\label{sumsigma1} 
   \begin{split}
     \sum_{\sigma}1  = 2^N\;, \qquad 
     \sum_{\sigma}\sigma_i = 0\;, \qquad 
     \sum_{\sigma} \sigma_i \sigma_j =  2^N \delta_{ij} \;, 
   \end{split} 
  \ee
 where in the last equation we used the Kronecker symbol $\delta_{ij}$ that is equal to $1$ for $i=j$ and zero otherwise.  
The second relation follows because we can perform the shift of summation variables $\sigma_i\rightarrow -\sigma_i$ 
to show that the sum goes to minus itself and is hence zero. 
The third relation follows similarly because for $i=j$ it reduces to the first relation due to  $\sigma_i^2=1$, 
and for $i\neq j$ we can perform again 
the index shift $\sigma_i\rightarrow -\sigma_i$. 

We can generalize the above relation iteratively by noting that by the same arguments the sum over all odd powers of $\sigma$ vanish, 
and for all even powers the result is non-zero (and then equal to $2^N$) if the $\sigma$'s pairwise have the same index. 
Unfortunately, it is not possible to write a tensor-type equation as above, since we have to distinguish cases. 
For instance,\footnote{Based on experience with Wick's theorem from QFT, at first one might be tempted to guess 
 \be
   \sum_{\sigma} \sigma_i\sigma_j \sigma_k\sigma_l =  2^N\big(\delta_{ij}\delta_{kl} +\delta_{ik}\delta_{jl} +\delta_{il}\delta_{jk} \big) \quad ? 
 \ee 
 but this would be incorrect for instance for $i=j=k=l$, where the right-hand produces $2^N\times 3$ but the sum on the left-hand side 
 is only $\sum_{\sigma} 1 = 2^N$. }   
 \be\label{sumsigma2} 
  \sum_{\sigma} \sigma_i\sigma_j \sigma_k\sigma_l = 
  \begin{cases}
      2^N \delta_{kl}   & \text{for $i=j$}\\
      2^N(\delta_{ik}\delta_{jl} + \delta_{il}\delta_{jk})  & \text{for $i\neq j$}
      \end{cases} 
 \;. 
 \ee
For the computation below we also need the expression for six powers of $\sigma$'s. For $i\neq j$, $k\neq l$ and $p\neq q$ we have 
 \be\label{sixsigmassssss} 
    \sum_{\sigma} \sigma_i\sigma_j \sigma_k\sigma_l \sigma_p \sigma_q = 2^N\big( 
    4\,\delta_{i(k}\,\delta_{l)(p} \,\delta_{q)j}  
    + 4\,\delta_{i(p}\,\delta_{q)(k} \,\delta_{l)j}\big) \;,   
 \ee
where we use the notation $a_{(i}b_{j)} := \frac{1}{2}(a_i b_j + a_j b_i)$ for symmetrizing indices. 
The combinatorics works as follows: the index $i$ needs to be paired with another index other than $j$, 
for which there are $4$ possibilities, after which the index $j$ needs to be paired with any of the 2 remaining 
indices in the group of $2$ with which $i$ was not paired, thus leaving $4\times 2=8$ possible terms. Writing out the symmetrizations (\ref{sixsigmassssss}) 
we have indeed $8$ terms. More generally, we have 
\be
 \sum_{\sigma} \sigma_{i_1}\sigma_{i_2}\cdots \sigma_{i_m}
 =
 \begin{cases}
  2^N\,, & \text{if each index appears an even number of times,}\\
  0\,, & \text{otherwise.}
 \end{cases}
\ee

\medskip

Using (\ref{sumsigma1}) and (\ref{sumsigma2}) in (\ref{partitionfunctionspinglasssss}), freely exchanging
the orders of (finite) summations,  and recalling $J_{ii}=0$ we obtain 
 \be
 \begin{split} 
  Z(\beta, J) &= \sum_{\sigma} 1 + \frac{\beta}{2}  \sum_{i,j} J_{ij}  \sum_{\sigma}  \sigma_i \sigma_j
  +\frac{\beta^2}{8} \sum_{i,j,k,l}J_{ij} J_{kl} \sum_{\sigma}   \sigma_i\sigma_j \sigma_k \sigma_l +\cdots \\
  &= 2^N\Big( 1+ \frac{\beta^2}{8}  \sum_{i,j,k,l}J_{ij} J_{kl}\big(\delta_{ik}\delta_{jl} +\delta_{il}\delta_{jk} \big)
  +\cdots  \Big) \\
   &= 2^N\Big( 1+ \frac{\beta^2}{4}  \sum_{k,l}J_{kl} J_{kl} 
  +\cdots  \Big)\,, 
 \end{split}  
 \ee
and hence 
 \be\label{oneoverZ} 
  \frac{1}{Z}  = \frac{1}{2^N}\Big( 1- \frac{\beta^2}{4}  \sum_{k,l}J_{kl} J_{kl} 
  +\cdots  \Big)\;. 
 \ee 
 
We can now compute the two-point function $\langle \sigma_i\sigma_j\rangle $, 
defined as in (\ref{2pointfunc}), (\ref{2poinsfuncr333}). 
To second order 
in $\beta$ we have for $i\neq j$: 
 \be
  \begin{split}
   \langle \sigma_i\sigma_j\rangle &= \frac{1}{Z} \sum_{\sigma} \sigma_i\sigma_j 
   \exp\Big(\tfrac{\beta}{2}  \sum_{k,l} J_{kl} \sigma_k \sigma_l\Big) \\
   &= \frac{1}{Z} \sum_{\sigma} \sigma_i\sigma_j 
   \Big(1+ \frac{\beta}{2}  \sum_{k,l} J_{kl} \sigma_k \sigma_l
   +\frac{\beta^2}{8} \sum_{k,l,p,q} J_{kl} J_{pq} \sigma_k\sigma_l\sigma_p \sigma_q +\cdots \Big) 
    \\
   &= \frac{1}{Z}
   \Big( \sum_{\sigma} \sigma_i\sigma_j + \frac{\beta}{2}  \sum_{k,l} J_{kl}  \sum_{\sigma} \sigma_i\sigma_j  \sigma_k \sigma_l
   +\frac{\beta^2}{8} \sum_{k,l,p,q} J_{kl} J_{pq}  \sum_{\sigma} \sigma_i\sigma_j  \sigma_k\sigma_l\sigma_p \sigma_q +\cdots \Big)  \\
     &= \frac{1}{Z}2^N
   \Big({\beta} J_{ij}    + {\beta^2} \sum_{k} J_{ik} J_{jk}   +\cdots \Big)\\
   &= {\beta} J_{ij}   +{\beta^2} \sum_{k} J_{ik} J_{jk}   +\cdots\;, 
  \end{split} 
 \ee
where from the fourth to the fifth line we used (\ref{sumsigma1}), (\ref{sumsigma2}) and (\ref{sixsigmassssss}), 
and in the last line we inserted (\ref{oneoverZ}).  
For $i=j$ we have, of course, $\langle \sigma_i\sigma_j\rangle = \langle \sigma_i^2\rangle =1$ and so in total: 
 \be
    \langle \sigma_i\sigma_j\rangle = \begin{cases}
       1 & \text{for $i=j$}\\
       {\beta} J_{ij}   +{\beta^2} \sum_{k} J_{ik} J_{jk}   +\cdots & \text{for $i\neq j$}
      \end{cases} \;. 
 \ee
There is a diagrammatic procedure to determine the perturbation theory that is closely analogous 
to the Feynman diagrams of QFT, in which each $J_{ij}$ is represented by an unoriented edge connecting
  vertices $i$ and $j$ \cite{Daboul,Friedli}. It should be noted, however, that in QFT $\beta$ plays the role of 
  the inverse of Planck's constant, $\beta\sim \frac{1}{\hbar}$, and so  the expansion here corresponds to 
  the `strong coupling' expansion in large $\hbar$.\footnote{See \cite{Douglas:2026ssj} for a brief 
  discussion of the significance of the strong coupling expansion  in the lattice field theory 
  formulation of Yang-Mills gauge theory.}

\newpage 
   
 \section{The Renormalization Group} 
 
 The renormalization group (RG) provides techniques that allow one to identify phase transitions without having 
 to compute the exact partition function. The rough picture  is as follows: Starting from a lattice with $N$ points and a microscopically 
 small lattice spacing we may 
 reasonably expect that the physics is essentially equivalent if we increase the lattice spacing somewhat  and sum over 
 the spins arising in all of the newly obtained  blocks (`blockspin'). This leads to a rewriting of the partition function 
 in terms of the remaining spins  that takes the same  Boltzmann form, with an energy function of the same 
 functional form, but with transformed coupling constants. This transition from the original coupling constants to the new 
 coupling constants  defines the RG transformation, which can be iterated. A fixed point of these RG transformations indicates
 the presence of a phase transition. 
 
 \subsection{RG of the  1d Ising model}
 
 We illustrate the idea with the (periodic) 1d Ising model (\ref{B=0Isung}) with partition function 
  \be\label{IsingPartRG} 
   Z_N = \sum_{\{\sigma_i=\pm 1\}} e^{-\beta E({\sigma})} = 
    \sum_{\{\sigma_i=\pm 1\}} e^{K\sum_{i}\sigma_i\sigma_{i+1} }\;, 
  \ee
 where we defined the rescaled couplings constant 
  \be
   K:=\beta J\;.
  \ee 
We will now partially perform the sum in (\ref{IsingPartRG}), the sum over all $\sigma_i$ with $i$ even, 
thereby effectively doubling the lattice spacing. 
In terms of (\ref{effectiveEnergyyyy}), (\ref{Peff}) above we determine the marginalized Boltzmann distribution 
obtained by integrating out $\sigma_2, \sigma_4, \ldots$,  leading to an effective energy function for $\sigma_1, \sigma_3, \ldots$.

In order to compute this, we consider the first such sum in (\ref{IsingPartRG}): 
 \be
  \sum_{\sigma_2=\pm 1} e^{K (\sigma_1\sigma_2+\sigma_2\sigma_3)} =2\cosh(K(\sigma_1+\sigma_3))\;. 
 \ee 
The goal is to bring this to the same functional form $e^{-\beta E'}$, with $E'$ the energy function of an  Ising model. 
This is almost possible, except for a constant term that needs to be included: 
 \be
  2\cosh(K(\sigma_1+\sigma_3)) = e^{2g+K'\sigma_1\sigma_3}\;. 
 \ee 
Note that on the new lattice $1$ and $3$ are nearest neighbors.  
The coefficients are fixed by taking  $\sigma_1=-\sigma_3$, for which $2\cosh(0)=2=e^{2g-K'}$ 
and comparing with $\sigma_1=\sigma_3$, for which $2\cosh(2K)=e^{2g+K'}$. 
These two equations are solved by 
 \be\label{RGStepISing} 
  K' = {\cal R}(K) := \frac{1}{2}\ln \cosh(2K) \;, \qquad g(K') =\frac{1}{2}(\ln 2+K')\;. 
 \ee 
  
The partition function then reduces to the sum over all remaining $\frac{N}{2}$ summation variables $\sigma_1, \sigma_3,\sigma_5,\ldots$ 
with an odd index: 
 \be\label{RGedIsingSum} 
  Z_N(K)  = \sum_{\sigma_1=\pm 1}\sum_{\sigma_3=\pm 1}\cdots \sum_{\sigma_{N-1}=\pm 1} 
  \exp\Big({\frac{N}{2}2g + K' \sum_{i}\sigma_i \sigma_{i+2}}\Big)\;, 
 \ee 
where we used that we performed $\frac{N}{2}$ sums (assuming $N$ is even).  
Defining the partition function of the 1d Ising model for any number $N$ of spin degrees of freedom by (\ref{IsingPartRG}) 
the above can be written as 
 \be
  Z_N(K) = e^{Ng(K')} Z_{\frac{N}{2}}(K')\;, 
 \ee 
where $K'$ is given by (\ref{RGStepISing}). 

Now we can iterate the RG steps any number of times. 
Calling this number $k$ and assuming  that $N=2^m$, where $m>k$, 
we can keep performing sums over half of the remaining degrees of freedom:  
 \be\label{iteratedRGs} 
 \begin{split}
  Z_N(K) &= e^{Ng(K')} Z_{\frac{N}{2}}(K') \\[1ex] 
  &= e^{Ng(K')}e^{\frac{N}{2}g(K'')} Z_{\frac{N}{2^2}}(K'') \\
   &
   = \exp\Big[ N\sum_{\ell=1}^{k} \frac{1}{2^{\ell-1}} g(K^{(\ell)})  \Big]\;   Z_{\frac{N}{2^k}}(K^{(k)}) \;, 
    \end{split} 
 \ee
where with (\ref{RGStepISing}) 
 \be\label{1DIsingRGflow} 
  K^{(\ell+1)} = {\cal R}(K^{(\ell)})  =  \frac{1}{2} \ln \cosh(2K^{(\ell)}) \;, \qquad
  g(K^{(\ell)}) = \frac{1}{2}\ln 2+\frac{1}{2} K^{(\ell)} \;. 
 \ee

 \bigskip 
 
\begin{figure}
    \centering
    \includegraphics[width=0.75\textwidth]{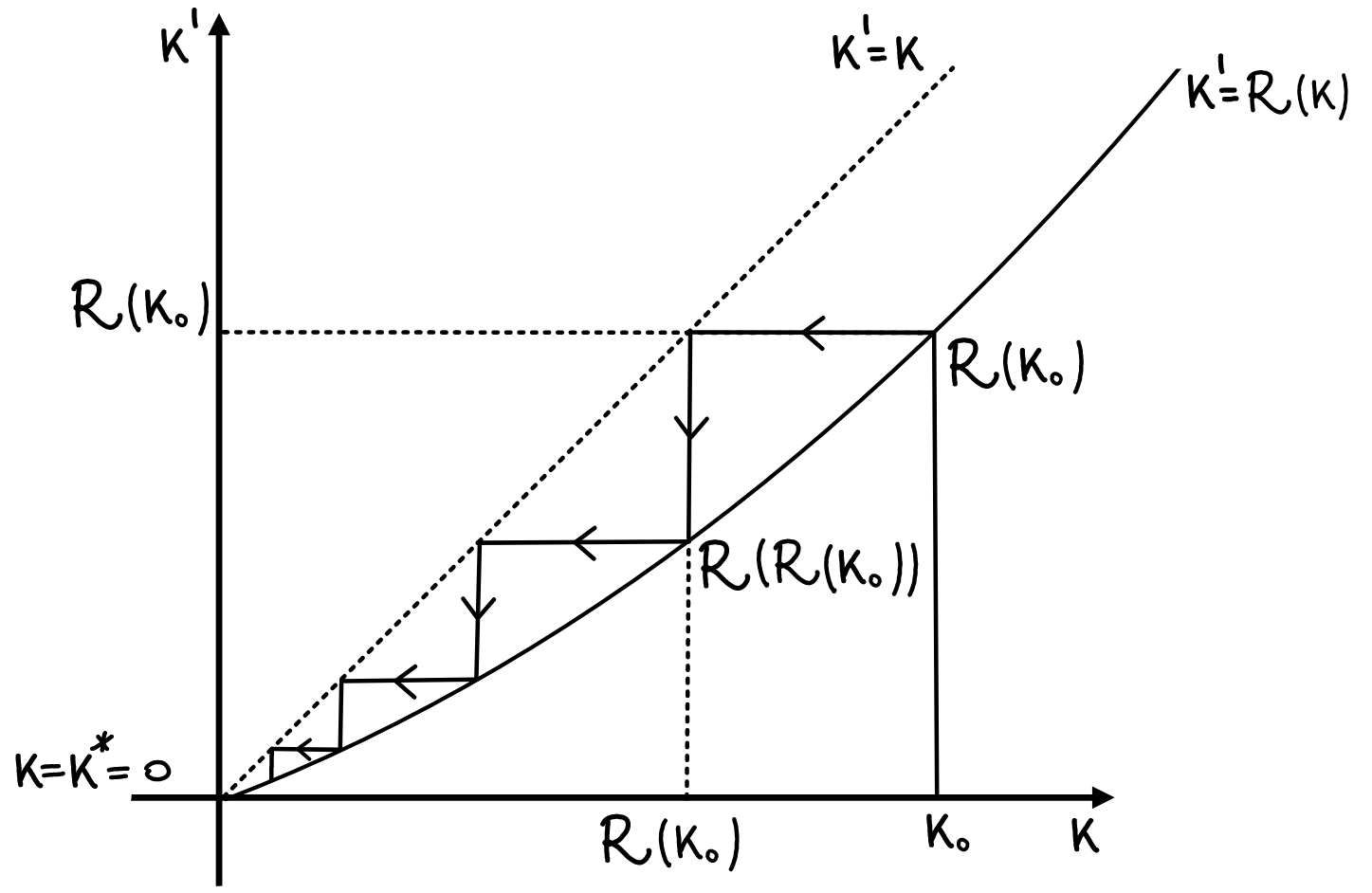}
    \caption{RG flow  of 1d Ising model  }
    \label{1DIsingRG}
\end{figure}

 \noindent\textit{RG fixed point:}\\[0.5ex]   
The equations (\ref{1DIsingRGflow}) can be viewed as evolution equations for a discrete dynamical system. One starts with some  
initial condition $K^{(0)}=K_0$ and then iteratively computes  $K^{(1)}$, $K^{(2)}$, etc., via (\ref{1DIsingRGflow}). (Since $g$ is 
determined by $K$ we can focus on the evolution of $K$.) This defines the (discrete) RG flow, 
which may or may not have a fixed point: a point that is left invariant by the evolution. 
As we will argue below, the presence of a non-trivial fixed  point of these RG transformations indicates a phase transition. 
A fixed point $K=K^*$ must obey 
 \be
  K^* = {\cal R}(K^*) = \frac{1}{2} \ln\cosh(2K^*)\;. 
 \ee
The only finite  solution is 
 \be
  K^*=0\;, 
 \ee 
in which the effective coupling constant vanishes. 
Keeping $J$ fixed this amounts to $\beta\rightarrow 0$ or $T\rightarrow \infty$. 
It is easy to see graphically, see  figure \ref{1DIsingRG},  that for an arbitrary starting  point $K_0$
the RG transformations  drive the coupling to zero.

Thus, there is no non-trivial fixed  point and hence no phase transition.

  \medskip 
 
  \noindent\textit{RG derivation of free energy density:}\\[0.5ex]   
It is tempting  to try to recover the exact free energy density (\ref{exactfreeenergylimit})
by taking  the simultaneous limits $N\rightarrow \infty$ and $k\rightarrow \infty$, i.e., taking the thermodynamic limit 
of infinite volume and sending the number of RG steps to infinity. In order to perform the sum $\sum_{\ell}$ in (\ref{iteratedRGs}), 
however, we would need a  closed form expression for $K^{(\ell)}$, which we do  not have. 
We can, however, recover (\ref{exactfreeenergylimit}) by taking the $N\rightarrow \infty$ limit after just one RG step. 
To this end we note that the first line of (\ref{iteratedRGs}) can be written as 
 \be
  Z_N(K) = (A(K))^{\frac{N}{2}} Z_{\frac{N}{2}}(K') \;, \qquad A(K):=2(\cosh(2K))^{\frac{1}{2}}\;. 
 \ee
From this it follows 
 \be
  \frac{1}{N} \ln Z_N(K) = \frac{1}{2} \ln A(K) + \frac{1}{2} \frac{1}{\frac{N}{2}} \ln Z_{\frac{N}{2}}(K') \;.
 \ee
Now taking the limit we obtain 
 \be
  \phi(K) := \lim_{N\rightarrow\infty} \frac{1}{N} \ln Z_N(K) = \frac{1}{2} \ln A(K)  + \frac{1}{2} \phi(K')\;, 
 \ee
or, recalling (\ref{RGStepISing}), 
 \be
    \phi(K) = \frac{1}{2} \ln A(K)  + \frac{1}{2} \phi\big(\tfrac{1}{2}\ln \cosh(2K) \big)\;. 
 \ee
This is a functional equation for $\phi(K)$. It is straightforward, although not entirely trivial, to verify that 
 the ansatz 
  \be
   \phi(K) = \ln\big(2\cosh(K)\big) 
  \ee 
 indeed solves this equation, in agreement with (\ref{exactfreeenergylimit}).

  \subsection{RG of the  2d Ising model}
 
\begin{figure}
    \centering
    \includegraphics[width=0.75\textwidth]{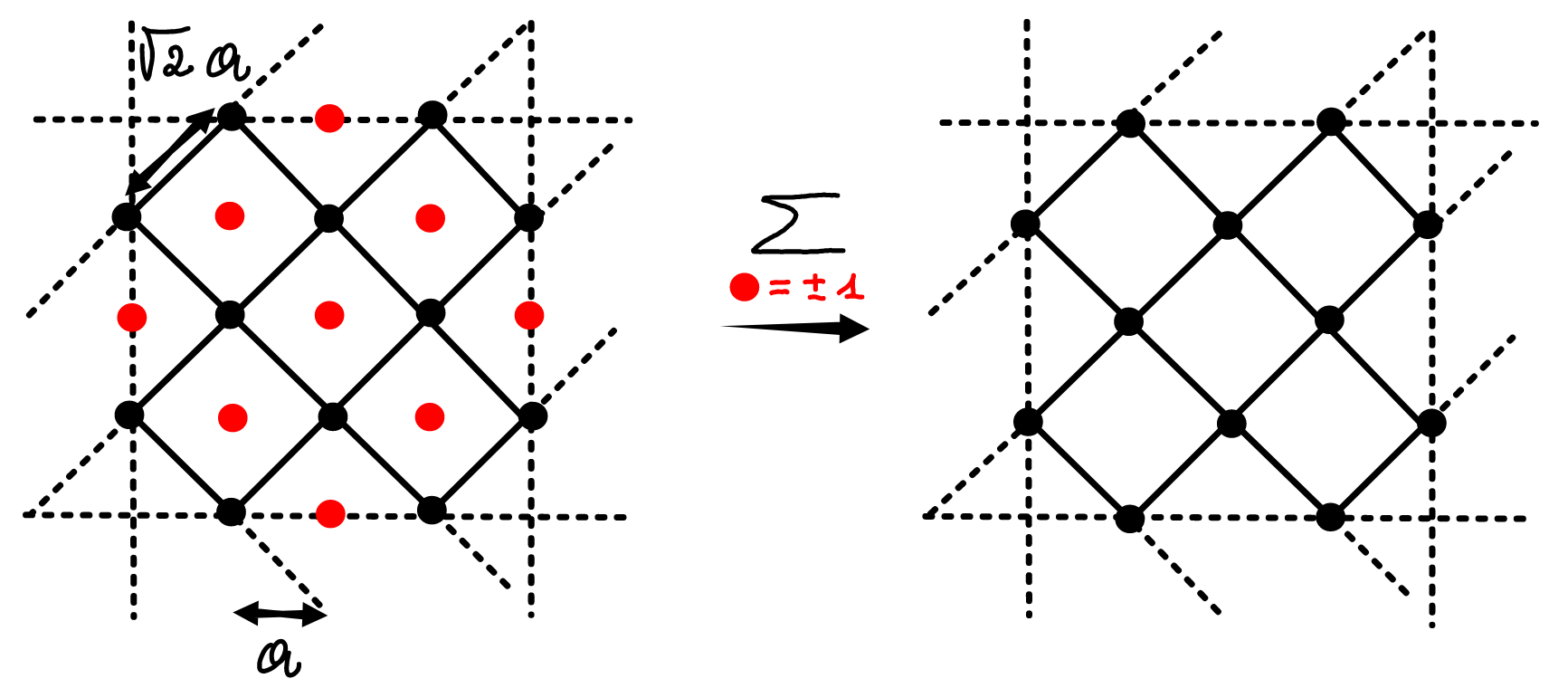}
    \caption{Integrating out spins of 2d Ising model  }
    \label{2Dlattice}
\end{figure}

 We now aim to apply the RG techniques to the $d=2$ Ising model with partition function 
  \be
   Z = \sum_{\{\sigma_i=\pm 1\}} e^{K\sum_{(ij)}  \sigma_i \sigma_j} \;, 
  \ee
 where $i=1,\ldots,N$ labels the  lattice points, and $\sum_{(ij)}$ is the sum over nearest-neighbors in 
 the square lattice. As displayed  in figure \ref{2Dlattice} of a square lattice, 
 we can sum over (integrate out) half of the spin degrees of freedom, which leads to a new 
 square lattice that has been rotated 
 by $45^{\circ}$ and whose lattice spacing $a$ has been increased by the factor $b= \sqrt{2}$. 
 
 Summing over one of the   $\sigma$ that are displayed in red in figure \ref{2Dlattice}  
 one  obtains an effective contribution for its four nearest neighbors:
  \be\label{typicalSUm} 
  \begin{split} 
   \sum_{\sigma=\pm 1} e^{K\sigma(\sigma_1+\sigma_2+\sigma_3+\sigma_4)} &= 2\cosh(K(\sigma_1+\sigma_2+\sigma_3+\sigma_4)) \\
   &= e^{A'+ \frac{1}{2}K'(\sigma_1\sigma_2+\sigma_1\sigma_4+\sigma_2\sigma_3+\sigma_3\sigma_4)
   +L'(\sigma_1\sigma_3+\sigma_2\sigma_4) + M'\sigma_1\sigma_2\sigma_3\sigma_4}\;, 
  \end{split} 
  \ee
where in the second line we indicated the resulting contributions to the new energy function 
after the RG step.\footnote{The factor of $\frac{1}{2}$ in the new nearest-neighbor terms with coupling $K'$ 
is there because contributions like $\sigma_1\sigma_2$ will be produced from \textit{two} summations, see  the figure. }
This requires a next-to-nearest neighbor interaction but also a quartic interaction. 
Without these, there is no solution for the new primed coefficients in terms of $K$, but including 
all four $A', K', L', M'$ we can solve the above equations. By considering all possible combinations of parallel and anti-parallel 
$\sigma_1,\ldots, \sigma_4$, which we leave as an exercise,  one finds: 
 \be
  \begin{split}
   A'(K) &= \ln 2+ \frac{1}{8} \ln \cosh(4K) + \frac{1}{2} \ln \cosh(2K) \;, \\
   K'(K) &= \frac{1}{4} \ln \cosh(4K)\;, \\
   L'(K) &= \frac{1}{8} \ln \cosh(4K) \;, \\
   M'(K) &= \frac{1}{8} \ln \cosh(4K) -\frac{1}{2} \ln \cosh(2K)\;. 
  \end{split} 
 \ee
Now, if we were to perform a second RG step presumably we would generated yet higher interactions, between 
next-to-next-to-nearest neighbors, etc. Thus, already for the 2d Ising model we cannot hope to determine the RG flow exactly. 

Rather, we will truncate to the $(K,L)$ subspace, and also Taylor expand to second order, 
 \be\label{lncoshexpansion} 
  \ln \cosh(x) = \ln \big(1+\frac{1}{2}x^2+\cdots\big)  \simeq \frac{1}{2}x^2\;, 
 \ee
which yields 
 \be\label{prelimRG} 
  K' = 2K^2\;, \qquad L' = K^2\;. 
 \ee  
These equations are too trivial to exhibit an RG fixed point, but actually they were too naive to begin with since we started 
with the basic Ising model without next-to-nearest neighbor interactions, then generated such an interaction (with coupling $L'$) 
and decided to keep it in the truncation. Thus, we should go back, include it also in the starting model (with couplings $L$) 
and see what happens after the first RG step. In a typical sum such as (\ref{typicalSUm}) we would then include the 
corresponding extra terms:
 \be 
  \sum_{\sigma=\pm 1} e^{K\sigma(\sigma_1+\sigma_2+\sigma_3+\sigma_4)+L(\sigma_1\sigma_2+\sigma_1\sigma_4
  +\sigma_3\sigma_4+ \sigma_2\sigma_3 )} \;. 
 \ee 
Note that the next-to-nearest neighbor interactions do \textit{not} include terms with $\sigma_1\sigma_3$ and $\sigma_2\sigma_4$ 
as $1$ and $3$, for instance,  have a larger Euclidean distance than $1$ and $2$.\footnote{This would not be the case in the taxi-cap geometry, where only the number of steps along an edge matter, but at least for a physical magnet the Euclidean distance is the reasonable one.}
The extra term does not involve the summation variable $\sigma$ and hence just factors  out and shifts $K'$ by $L$ (taking into account the factor of $2$ elaborated on in footnote 4), 
so that (\ref{prelimRG}) is replaced by 
 \be\label{d2IsingRG} 
 \begin{split} 
  K' &= 2K^2 + L \;, \\
   L' &= K^2\;. 
\end{split} 
 \ee
The iterations of these equations define the truncated RG flow of the 2d Ising model.

 \noindent\textit{RG fixed point and stability:}\\[0.5ex]   
An RG fixed point  $(K^*, L^*)$ of (\ref{d2IsingRG})  needs to obey 
 \be\label{fixedpoint} 
  \begin{split} 
  K^* &= 2(K^*)^2 + L^* \;, \\
   L^* &= (K^*)^2\;, 
\end{split} 
 \ee 
which are quickly seen to admit, apart from the trivial solution $(K^*, L^*)=(0,0)$, the non-trivial solution 
 \be\label{nontrivialFix} 
  (K^*, L^*)=\Big(\frac{1}{3}\,, \;\frac{1}{9}\Big)\;. 
 \ee
 
\begin{figure}
    \centering
    \includegraphics[width=0.50\textwidth]{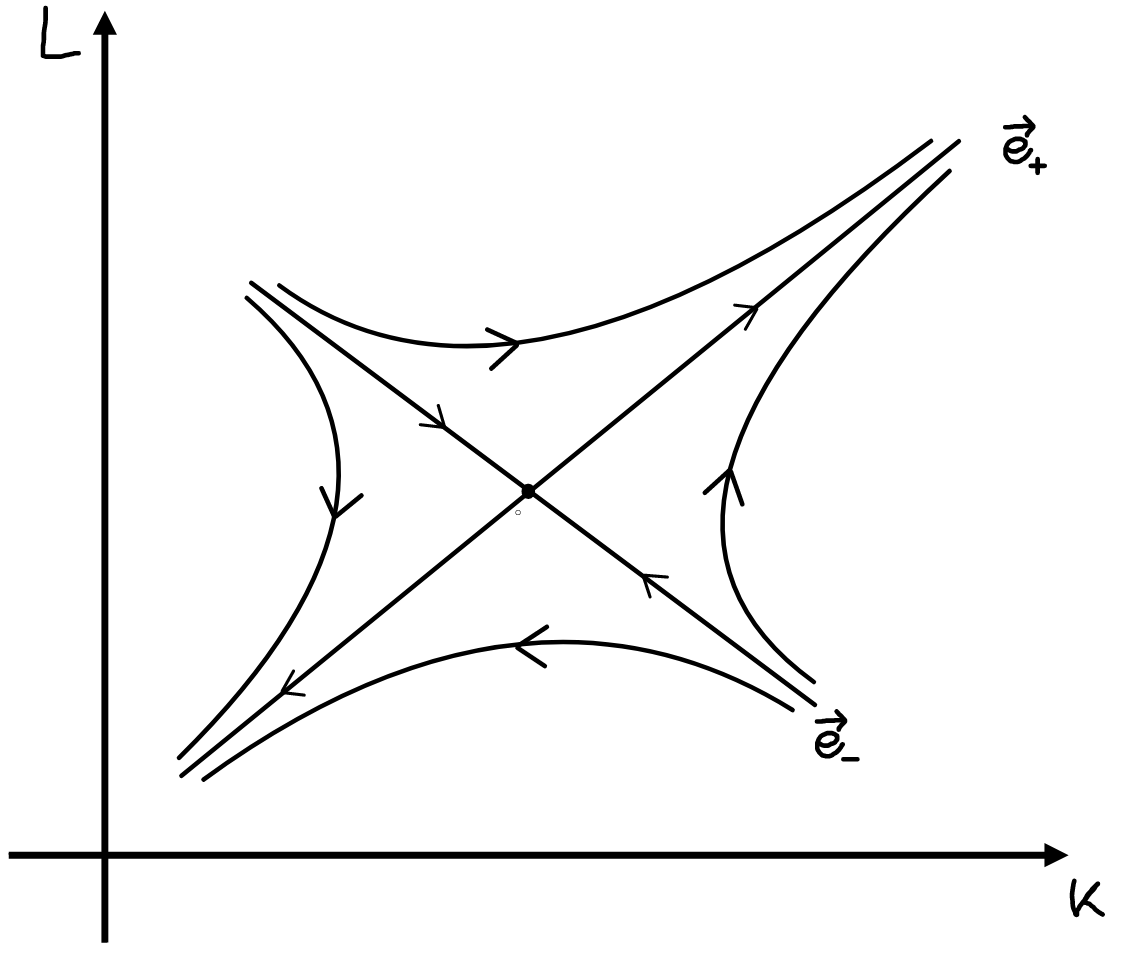}
    \caption{RG flow diagram of 2d Ising model  }
    \label{IsingFlow}
\end{figure}

We next inspect the stability of this fixed point by linearizing the RG equations  (\ref{d2IsingRG}), 
writing 
 \be
  K= K^*+\delta K\;, \qquad L = L^*+\delta L\;, 
 \ee
where $|\delta K|\ll 1$ and $|\delta L|\ll 1$ are small perturbations.  
Inserting this into (\ref{d2IsingRG}), 
 \be
 \begin{split} 
  K^*+(\delta K)' &= 2(K^*+\delta K)^2 + L^*+\delta L \;, \\
  L^*+(\delta L)'&= (K^*+\delta K)^2\;, 
\end{split} 
 \ee
expanding to first order in perturbations 
and using (\ref{fixedpoint}) one obtains 
 \be
 \begin{split} 
  (\delta K)' &= 4K^*\delta K + \delta L\;,  \\
  (\delta L)' &= 2K^*\delta K \;. 
 \end{split}  
 \ee
Specializing to the fixed point (\ref{nontrivialFix}) and using matrix notation this becomes 
 \be\label{matrixRGstep} 
  \begin{pmatrix} \delta K' \\ \delta L' \end{pmatrix} = 
  \begin{pmatrix} \frac{4}{3} && 1 \\ \frac{2}{3} && 0 \end{pmatrix}  \begin{pmatrix} \delta K \\ \delta L \end{pmatrix}\;. 
 \ee 
 Diagonalizing the 2-by-2 matrix yields the eigenvalues 
  \be
   \lambda_{\pm} = \frac{1}{3} (2\pm \sqrt{10}) 
   =\begin{cases}
      1.72 \\
      -0.39
      \end{cases}    \;, 
  \ee
and thus $|\lambda_+|>1$ and $|\lambda_-|<1$. This distinction is of significance since expanding into eigenvectors $\vec{e}_{\pm}$, 
 \be
    \begin{pmatrix} \delta K \\ \delta L \end{pmatrix} = c_+ \vec{e}_{+} + c_- \vec{e}_-\;, 
 \ee
and applying the RG steps  (\ref{matrixRGstep})  $\ell$  times yields 
 \be
   \begin{pmatrix} \delta K^{(\ell)}  \\ \delta L^{(\ell)} \end{pmatrix} = c_+ (\lambda_+)^{\ell} \vec{e}_{+} + c_-(\lambda_-)^{\ell}  \vec{e}_-\;. 
 \ee
Because of  $|\lambda_+|>1$ any perturbation along $\vec{e}_+$ becomes larger and larger, hence indicating an unstable direction 
(or a \textit{relevant} perturbation), while due to $|\lambda_-|<1$ any perturbation along $\vec{e}_-$ becomes smaller and smaller, 
thus being driven back to the RG fixed point, indicating a stable direction (or an \textit{irrelevant} perturbation).   
See the flow diagram in figure \ref{IsingFlow}.

Here is the core statement relating RG fixed points to phase transitions: \\
\textit{ A system exhibits a second-order phase transition 
if it has a non-trivial RG fixed point with a relevant perturbation.}   
Thus, in the parameter space of coupling constants, there is at least  one unstable direction, 
which has to be fine-tuned to reach the fixed-point. 
Temperature is one such parameter that has to be fine-tuned to reach the critical temperature of the phase transition. 
Moreover, we will see that at RG fixed points the system becomes scale invariant. 

Thus, the above result shows the presence of a second-order phase transition. Using  $K_c= K^*=\frac{1}{3}$ 
and returning to the original variables we have 
 \be\label{critTEMP} 
  \frac{J}{kT_c} = \frac{1}{3} \qquad \Leftrightarrow \qquad kT_c = 3J\;. 
 \ee
There is also an exact solution of the 2d Ising model due to Onsager that rigorously establishes 
the second order phase transition at the critical temperature $kT_c = 2.27\,J$ 
 \cite{Onsager:1943jn}. 
Given the crude approximations of the RG analysis,  the above result (\ref{critTEMP}) 
is pleasantly close to the exact result.

 \subsection{Generalities and Scaling Laws}\label{scalinglaws}  
 
 Based on our experience with two explicit examples of RG transformations we will now make some general remarks to explain the significance of RG fixed points for phase transitions. (See \cite{Goldenfeld} for an introduction.) 
 We still assume to have only finitely many spin degrees of freedom $\sigma_i=\pm 1$, where $i=1,\ldots, N=|\Lambda |$ labels 
 the points of a cubic lattice $\Lambda\subset \mathbb{Z}^d$. We assume that we have an energy function 
  \be
   {\cal H}_{\Lambda}({\sigma};{g}) := -\beta E_{\Lambda}({\sigma};{g}) = \sum_I g_I {\cal O}_I({\sigma}) \;, 
  \ee
 where for convenience we rescaled by $-\beta$, so that also the coupling constants
  \be
   {g}=(g_I)= (g_1,g_2,\ldots) 
  \ee
should be thought of as being rescaled by $-\beta$. The `operators'\footnote{This is a terrible yet standard name.} or observables 
 ${\cal O}$ are functions of the $\sigma_i$. For instance, for the class of  models suggested by the 2d Ising model after performing one 
 RG step, c.f..~(\ref{typicalSUm}), the coupling constants are ${g}=(K,L,M,\ldots)$ and the corresponding 
 operators ${\cal O}_1=\sigma_1\sigma_2+\sigma_1\sigma_4+\sigma_2\sigma_3+\sigma_3\sigma_4$, etc.  
 We assume that we have included all possible operators 
 that may be generated by an RG transformation.
 
 Next, we split the spin degrees of freedom arbitrarily as 
  \be
   {\sigma}  = ({\sigma}', {\tau})\;, 
  \ee 
with the goal to  perform the partial sum in the partition function over the ${\tau}$.  
This defines a transformed energy function: 
  \be\label{transfH} 
   \exp\big({\cal H}_{{\Lambda}/{b}}({\sigma}', {g}') + \Delta_{\Lambda}(g)\big)  :=
   \sum_{{\tau}} \exp\big({\cal H}_{\Lambda}({\sigma}', {\tau}; {g}) \big) \;, 
  \ee
 where by $\Lambda/b$ we denote the cubic lattice where all side length have been scaled by $b>1$.  
 In the above, we split off $\Delta_{\Lambda}(g)$, which is characterized by not depending on the $\sigma$. 
 The non-trivial statement implicit in the above is that the same universal function ${\cal H}_{\Lambda}$ governs 
 the result after integrating out the $\tau$, just with respect to the scaled lattice $\Lambda/b$ and with transformed 
 coupling constants 
  \be\label{RGFLOW} 
   {g}' = R({g}) \qquad \Leftrightarrow \qquad g_I' = R_{I}(g)\;, 
  \ee
 for some nonlinear transformations $R$. This is justified by our above assumption that the set of 
 operators ${\cal O}_I$ is complete in that no new operators are generated  in any RG step. 
 The partition function can then be rewritten in terms of (\ref{transfH}): 
  \be
  \begin{split} 
   Z_{\Lambda}({g}) &= \sum_{{\sigma}} e^{{\cal H}_{\Lambda}({\sigma}; {g})} 
   = \sum_{{\sigma}',{\tau}} e^{{\cal H}_{\Lambda}({\sigma}', {\tau}; {g})} \\
   &=\sum_{{\sigma}'}  \exp\big({\cal H}_{{\Lambda}/{b}}({\sigma}', {g}') + \Delta_{\Lambda}(g)\big) \\
   &= e^{\Delta_{\Lambda}(g)} Z_{\Lambda/b}({g}')\;. 
  \end{split} 
  \ee
  
 We will next assume that  the RG flow equations (\ref{RGFLOW}) have a non-trivial fixed point ${g}^*$, satisfying 
  \be\label{RGFIxed} 
   {g}^* = R({g}^*)\;, 
  \ee 
 and explore its consequences.  We begin by linearizing the RG equations around this fixed point: 
  \be
   {g} = {g}^*+\delta{g}\;. 
  \ee
 Expanding (\ref{RGFLOW}) to first order in the perturbation we have 
  \be
   {g}^* + (\delta {g})' = R({g}^*) + \mathfrak{L}(\delta {g}) +\cdots \;, 
  \ee 
 where $\mathfrak{L}$ is the linear operator obtained by linearizing $R$. 
 Recalling (\ref{RGFIxed}) this implies 
  \be\label{inftRGStep} 
   (\delta {g})' =\mathfrak{L}(\delta {g}) \;. 
  \ee
 This equation defines the RG step of the perturbation $\delta {g}= {g}  - {g}^*$, 
 which by construction is a linear transformation. 
 
 We now make the assumption (that tends to be obeyed in practice) that the linear operator $\mathfrak{L}$ is 
 diagonalizable, with eigenvalues $\lambda_I$. 
 Performing then a change of basis $\delta g_I\rightarrow \Lambda_I{}^{J} \delta g_{J}$ we can assume that 
 the  $\delta g_I$ are in an eigenbasis so that  (\ref{inftRGStep}) becomes 
  \be\label{RGEigenSTEP} 
   (\delta g_I)' = \lambda_I \,\delta g_I\;. 
  \ee
 It is customary to rewrite this by defining 
  \be\label{gammaredef} 
   y_I  := \frac{\ln |\lambda_I|}{\ln b} \qquad \Leftrightarrow \qquad b^{y_I} = |\lambda_I|\;, 
  \ee
 where $b>1$ is arbitrary in principle but here taken to be the scale factor by which one rescales the lattice. 
 In terms of this, (\ref{RGEigenSTEP}) becomes 
  \be\label{bscaledRGstep} 
    (\delta g_I)' = \pm\, b^{y_I} \,\delta g_I\;. 
  \ee
 Iterating this RG transformation $\ell$ times we have 
  \be
   (\delta g_I)^{(\ell)} = \pm\, b^{y_I\ell } \,\delta g_I\;. 
  \ee 
 Whether the fixed point is stable, i.e., whether the perturbations become smaller or larger, depends on whether $y_I<0$
 or $y_I>0$, suggesting the following classification: 
  \be
   \begin{split}
    &|\lambda_I|>1\;, \;\; y_I > 0\,: \qquad \text{\textit{relevant} perturbation } \\
    &|\lambda_I|<1\;, \;\; y_I < 0\,: \qquad \text{\textit{irrelevant} perturbation } \\
    &|\lambda_I|=1\;, \;\; y_I = 0\,: \qquad \text{\textit{marginal } perturbation } 
   \end{split}
  \ee

\bigskip

\noindent\textit{Scaling laws:}\\[0.5ex]  
We will now derive from the existence of an RG fixed point with a relevant perturbation 
so-called scaling laws that are characteristic for phase transitions or critical phenomena. 
To this end we employ (\ref{bscaledRGstep}) and rewrite it  as 
 \be
  (g_I - g_I^*)' = g'_I - g_I^*=  \pm\, b^{y_I  }(g_I - g_I^*)\;. 
 \ee 
We specialize this to the coupling constant $\beta$ or temperature $T$ itself. (All coupling constants such as $J$ of the Ising model 
were scaled by $\beta$ and so any one of them can be viewed as temperature with the original coupling entering  the energy function held 
fixed.) We then have 
 \be\label{TTRGTRANS} 
  (T-T^*)' = \pm b^{y_t} (T-T^*)\;, 
 \ee
where we denote by $y_t$ (defined as in (\ref{gammaredef})) the factor corresponding to temperature. 
We can make the notation slightly less cumbersome by defining 
 \be\label{DEFt} 
  t:= \frac{T^*-T}{T^*} = 1-\frac{T}{T^*} \;, 
 \ee
which by (\ref{TTRGTRANS}), and using that $T^*$ is invariant, RG transforms as 
 \be
  t' = \pm \,b^{y_t} \,t\;. 
 \ee  
Iterating this $n$ times we have 
 \be\label{ntimesiteratedRGoft} 
  t^{(n)} =  \pm \,b^{y_t n } \,t\;. 
 \ee  
We assume that $t$ is a relevant perturbation, i.e., $y_t>0$.

Our goal is to explore the behavior of the system as the temperature $T$ is increased from zero to the 
critical temperature $T^*$, for which we let $t$ vary in the interval 
 \be\label{tInterval} 
  1\ \geq \ t \ \geq \ 0 \;. 
 \ee  
We will then tune the number of RG steps and follow 
the \textit{correlation length} $\xi(T)$, which 
can be defined via the expectation value (two-point function) 
\be\label{TwoPointFunctionCorr} 
 \langle \sigma_i \sigma_j\rangle \ = \ \exp\Big({-\frac{|i-j|}{\xi(T)}} \Big) \qquad \text{for $N\rightarrow \infty$}
 \;. 
\ee  
The correlation length  is a measure for critical phenomena: it  goes to  
infinity at the critical temperature. 
Since in our conventions the $i, j$ in (\ref{TwoPointFunctionCorr}) 
are integers indicating the lattice positions, the correlation length $\xi$ is a dimensionless 
number. If we assign the actual lattice spacing a concrete length scale such as  $a=1 \mathring{A}=10^{-10}m$  
 we have to view the  
\textit{physical} correlation length as $\xi$ times this fundamental length scale, 
i.e.
 \be
  \xi_{\rm phys} = \xi a\;. 
 \ee 
  Now, let us view  $\xi$ as a function of  (\ref{DEFt}) and do one 
RG transformation from   $\xi=\xi(t)$ to $\xi'=\xi(t') $.  The {physical} correlation length $\xi_{\rm phys}$ 
is unchanged, but we have to remember 
that the new $\xi'$ measures this length in terms of the lattice spacing $a'$ 
that has been increased by the factor $b$, i.e., $\xi_{\rm phys}=\xi' a' $, where  $a'=ab$.  Thus, 
 \be
  \xi = b\xi' \qquad \Leftrightarrow \qquad \xi' = \frac{\xi}{b}\;. 
 \ee
Iterating the RG transformation $n$ times we have 
 \be
  \xi(t) = b \,\xi(t') = b^n\, \xi(t^{(n)}) 
 \ee 
 and thus with (\ref{ntimesiteratedRGoft}) 
  \be\label{transforlawxi} 
   \xi(t) = b^n\,  \xi(\pm \,b^{y_t n} t) \;. 
  \ee
  
 In order to derive the scaling laws we choose the number $n$ of RG transformations as 
  \be\label{numberRG} 
   n\simeq \frac{1}{y_t}\Big(1-\frac{\ln t}{\ln b} \Big) \;, 
  \ee 
by which we mean that we pick an integer close to the right-hand side. 
This is well-defined as an integer in the limit  thanks to (\ref{tInterval}):  as $t$ approaches the critical  value $t=0$ from above 
we have $-\ln t\rightarrow \infty$ and hence $n\rightarrow \infty$ (where we must also recall 
that $t$ is a relevant perturbation so that $y_t>0$).  
With this choice we have 
 \be
  y_t\, n \ln b \simeq  \ln b - \ln t = \ln \frac{b}{t} 
 \ee
and hence 
 \be\label{SFDFSGDG}
  \big(e^{\ln b}\big)^{y_t n} = b^{y_t n} = \frac{b}{t} \;. 
 \ee
This relation explains the significance of the choice of $n$ in (\ref{numberRG}),  for 
 using the last equality in (\ref{transforlawxi}) we have
 \be
   \xi(t) = b^n\,  \xi(\pm\, b) \;. 
 \ee 
Moreover, since (\ref{SFDFSGDG})  implies $b^{n}=\big(\frac{b}{t}\big)^{\frac{1}{y_t}}$, we have 
 \be
   \xi(t) = \Big(\frac{b}{t}\Big)^{\frac{1}{y_t}} \,  \xi(\pm\, b) \;. 
 \ee
Up to numerical coefficients we have thus established the temperature dependence or scaling law 
 \be
  \xi(t) \sim t^{-\frac{1}{y_t}} = t^{-\nu} =\frac{1}{t^{\nu}} \;, 
 \ee
where we defined  \textit{critical exponent}
 \be
  \nu:=\frac{1}{y_t}\;. 
 \ee
Returning to standard  temperatures this is the universal form of the correlation length near the 
critical temperature: 
 \be
  \xi(T) \sim 
  \Big(1-\frac{T}{T^*}\Big)^{-\nu}\;, 
 \ee 
which  we have thus derived from the existence of an RG fixed point with a relevant perturbation.

 \section{Boltzmann Machines} 
 
 In this section we define  Boltzmann machines, which were originally introduced  as models 
 for the neural network that exists in our brain and  subsequently proposed as an approach to 
artificial intelligence (AI). While Boltzmann machines are not directly used in the modern AI revolution, 
they inspired the techniques of deep learning, which is at the heart of modern AI.

\subsection{Neurons and Hopfield Networks}

The basic idea of a neuron is as a computational unit with $n$ inputs $x_1,\ldots, x_n$ and one output $y$ \cite{McCulloch:1943cnb}. 
The neuron carries certain weights, a vector $w_i$, $i=1, \ldots, n$, and a bias $b$, and performs the following computation: 
It takes the \textit{linear} combination $\sum_i w_i x_i$ and then adds the bias $b$, thus making it an affine  linear transformation. 
Subsequently, a \textit{nonlinear} so-called activation function is applied, which in the simplest cases is the step function $\Theta(x)$ that is 
zero for $x<0$ and one for $x>0$. Thus, the neuron computes from the inputs $x_i$ the nonlinear function:
 \be
 y = f_{\theta}(x) = \Theta\Big(\sum_iw_ix_i + b\Big) \;, 
 \ee  
where the notation $\theta=(w_i,b)$ is sometimes used for the totality of weights. 

It is easy to see that such a  neuron can perform simple logical operations. For instance, to define AND we 
take $x_1,x_2\in \{0,1\}$ and set $w_1=w_2=1$, $b=-1.5$. The above function then computes $y=1$ for $x_1=x_2=1$ 
and $y=0$ otherwise. Thus, interpreting $0$ as false and $1$ as true this is the AND operation: $y=x_1\wedge x_2$. 
Similarly, one may implement OR and NOT, but curiously not XOR (exclusive OR). The latter fact was taken as evidence that 
these and slightly more general models for neural networks (`perceptrons') were insufficient for the needs of AI \cite{Minsky}, 
but the modern development has shown that if one considers deep networks with several layers  the problems disappear. 

We now turn to the \textit{Hopfield}  network. It consists of any number $N$ of neurons (nodes of a graph) labelled by $i,j,\ldots=1,\ldots, N$, 
and one assigns $\sigma_i=\pm 1$ to each neuron. (We could also use $0$ and $1$, but $\pm 1$ is equally convenient and brings out the close relation to Ising and spin-glass models.) 
We assume that all neurons are mutually connected. The connection between any two neurons $i$ and $j$ is given by a symmetric weight: 
 \be\label{symmetricWeights} 
  w_{ij}=w_{ji}\;, \qquad w_{ii}=0\;, 
 \ee
with the second condition making explicit that each neuron may receive inputs from all other $N-1$ neurons, 
but not itself. In addition, each neuron carries a bias denoted by  $b_i$. 
The configuration of the network is given by an assignment of neuron states, e.g.~${\sigma}=(+1,\ldots, +1)$. 

Next we define a certain dynamics of this network by giving an update rule $\sigma_i\rightarrow \sigma_i'$ 
that changes the state of a given  neuron, or leaves 
it unchanged, depending on the inputs from all other neurons. We define 
 \be\label{aiiiii} 
  a_i:= \sum_{j} w_{ij}\sigma_j  + b_i
 \ee
and declare that the state of neuron $i$ is updated to 
 \be\label{UdateRUle} 
  \sigma_i ' = \begin{cases}
      +1  & \text{for $a_i\geq 0$}\\
      -1 & \text{for $a_i< 0$}
      \end{cases}  \;, 
 \ee 
independent of what the current state $\sigma_i$ is. Concretely, we can imagine that we run in definite time steps through 
the neurons of the network, one after another (for instance, in the canonical order  from $i=1$ to $i=N$ and then back to $i=1$), 
and update each neuron according to (\ref{UdateRUle}).

The above dynamical system has the interesting property that it decreases or leaves unchanged the function 
 \be\label{SpinGlassBoltzmann} 
  E({\sigma}) := -\frac{1}{2}\sum_{i,j} w_{ij} \sigma_i\sigma_j - \sum_i b_i \sigma_i \;, 
 \ee
which we recognize as the energy function (\ref{spinglass}) of spin-glass models in an external magnetic field $b_i$ (and after renaming 
the couplings $J_{ij}$ to $w_{ij}$). To prove this claim let us focus on the updating of one specific  neuron that, without loss 
of generality, we call $1$. Thus, the configuration change obeys  $\sigma_i'=\sigma_i$ for $i\neq 1$, and so the change 
in energy is 
 \be\label{energyDifference} 
 \begin{split} 
  E({\sigma}') - E({\sigma}) &= -\frac{1}{2}\sum_{i,j} w_{ij} \sigma'_i\sigma'_j - \sum_i b_i \sigma'_i 
  +\frac{1}{2}\sum_{i,j} w_{ij} \sigma_i\sigma_j + \sum_i b_i \sigma_i  \\
  &= -\sum_j w_{1j} \sigma_1'\sigma_j -b_1\sigma_1' + \sum_j w_{1j} \sigma_1\sigma_ j + b_1\sigma_1  \\
  &= -\sum_j w_{1j} (\sigma_1'-\sigma_1) \sigma_j -b_1(\sigma_1'-\sigma_1) \;, 
 \end{split} 
 \ee
where we used in the second line that all terms with $i,j\neq 1$ cancel and that, therefore, only the terms where either  
$i=1$ or $j=1$ contribute.  Now, in the update the neuron's state either changes, in which case $\sigma_1'=-\sigma_1$, or it doesn't, 
in which case $\sigma_1'=\sigma_1$. We thus have 
 \be\label{energyDifference2} 
   E({\sigma}') - E({\sigma}) =  \begin{cases}
      -2\big(\sum_j w_{1j} \sigma_j  + b_1\big) \sigma_1' & \text{for $\sigma_1'=-\sigma_1$}\\
      0 & \text{for $\sigma_1'=\sigma_1$}
      \end{cases} 
 \ee
Since by the update rule (\ref{UdateRUle}) we either have $a_1=\sum_j w_{1j} \sigma_j  + b_1\geq 0$ and $\sigma_1'=+1$ 
or $a_1<0$ and $\sigma_1'=-1<0$ it follows that 
 \be
  E({\sigma}') - E({\sigma})  \leq 0\;, 
 \ee
as we wanted to prove.

We have just learned that the dynamics (\ref{UdateRUle})  moves the configurations to lower and lower energy, 
so that the system eventually settles in a (local) energy minimum. This fact was the basis for Hopfields idea to model the 
memory functions of the brain in terms of the space of ground states of a spin glass system, 
see \cite{Quanta} for a popular account.\footnote{It should be emphasized, however, that the Ising spin type model for neurons 
with excitations $\pm 1$ is very far from the functioning of actual biological neurons. In particular, the assumption of symmetric weights in (\ref{symmetricWeights}) is unrealistic. See  \cite{Dayan,Gerstner} for an exposition of real biological networks.} 
Since, as we saw, spin glass models have 
a very rich space of ground states, we can imagine that each such state encodes a certain memory, which can then be remembered or evoked  
by flowing with the natural time evolution (\ref{UdateRUle}) to such a ground state.

As an illustration, let us consider a set of  neural configurations 
 \be\label{CONFig} 
  \{e_{i}{}^{A} \in \{\pm 1\} \,, \; i=1,\ldots, N\}  \;, 
 \ee  
where $A=1,\ldots, M$ labels the different configurations (that we can think of as specific memories such as pictures). 
To simplify the following discussion we assume $M<N$ and the  relation 
 \be\label{Orthogonal} 
   \frac{1}{N} \sum_i e_{i}{}^{A} e_{i}{}^{B} = \delta^{AB}\;, 
 \ee
implying that the neural configurations are orthogonal in this sense. 
Let us now define the weights of a Hopfield network as follows: 
 \be
  w_{ij} := \frac{1}{N} \sum_A e_{i}{}^{A} e_{j}{}^{A} \quad (i\neq j)\;, \qquad w_{ii}=0\;, \qquad b_i:=0\;.  
 \ee
The claim is that in this network the $M$ configurations (\ref{CONFig}) are all invariant under the time evolution
(are fixed points of the update rule (\ref{UdateRUle})) 
and therefore must correspond to local minima of the energy. 
To verify this, say the network is in the configuration that is number $14$ in the above list (\ref{CONFig}): 
 \be
  \sigma_i = e_i{}^{14} \;. 
 \ee
In order to apply the update rule (\ref{UdateRUle}) we compute next (\ref{aiiiii}): 
 \be
 \begin{split} 
  a_i &=\sum_j w_{ij} \sigma_j + b_i  = \frac{1}{N} \sum_{j\neq i} 
   \sum_A e_{i}{}^{A} e_{j}{}^{A} \sigma_j 
  =\frac{1}{N} \sum_{j\neq i}  \sum_A e_{i}{}^{A} e_{j}{}^{A} e_j{}^{14} \\
  &=  \sum_A e_{i}{}^{A}  \frac{1}{N} \big( \sum_{j}  e_{j}{}^{A} e_j{}^{14} -  e_{i}{}^{A} e_i{}^{14}\big)\\
  &=   \sum_A \Big(e_{i}{}^{A} \delta^{A,14}-\frac{1}{N} e_{i}{}^{14} \Big)  = \Big(1-\frac{M}{N}\Big) e_{i}{}^{14}  \;, 
 \end{split} 
 \ee
where we used (\ref{Orthogonal}) 
and $(e_i{}^{A})^2=1$ since $e_{i}{}^{A}=\pm 1$.  
Since we assumed $M<N$, $a_i$ has the same sign as $\sigma_i=e_{i}{}^{14} $. 
Thus, for $\sigma_i=+1$ we have $a_i>0$ and hence $\sigma_i'=+1$, 
and for $\sigma_i=-1$ we have $a_i<0$ and hence $\sigma_i'=-1$. Either way, the state of the neuron does not change, 
and hence we have a fixed point of the update rule.

The above shows that the neural network may encode a large number of fixed configurations as local minima 
of the spin-glass energy  and that the dynamics drives the system to one such minimum. 
The disadvantage of this deterministic evolution is that once such a configuration has been reached one cannot 
leave it again: the dynamics leaves the state invariant. This can be improved by adding a stochastic element governed 
by a `temperature' parameter to the dynamics, leading to the Boltzmann machines to which we turn now.

\subsection{Boltzmann Machines }

A Boltzmann machine is given by a Hopfield network as defined above but with a probabilistic time evolution. The state or 
configuration of the system is updated by the rule 
 \be\label{PROBUpdate} 
  \sigma_i ' = \begin{cases}
      +1  & \text{with probability $\sigma(2\beta a_i)$}\\
      -1 & \text{with probability $1-\sigma(2\beta a_i)$ }
      \end{cases}  \;, 
 \ee 
where $\beta$ is the temperature (or rather coolness) parameter, we recalled (\ref{aiiiii}), 
 \be\label{aiiiii22} 
  a_i:= \sum_{j} w_{ij}\sigma_j  + b_i\;, 
 \ee
and we introduced the important \textit{sigmoid function} or  \textit{logistic function}:\footnote{Sometimes 
sigmoid function refers to any S-shaped function, including $\tanh(x)$, while the logistic function refers to the 
specific function (\ref{sigmoid}).}
 \be\label{sigmoid} 
  \sigma(x):=\frac{1}{1+e^{-x}}\;. 
 \ee  
This function is popular because it is a smooth approximation to the step function, which is obtained in the zero temperature 
limit $\beta\rightarrow \infty$. Moreover, the logistic function  is convenient because its derivative can be written 
in terms of the original function as follows: 
 \be
  \sigma'(x) = \sigma(x) (1-\sigma(x)) \;. 
 \ee
Using (\ref{sigmoid}) in (\ref{PROBUpdate}) we can encode  the probabilistic update rule by saying that the 
probabilities for $\sigma_i'$ being $+1$ or $-1$ are given by 
 \be\label{stochupdaterules} 
  P(\sigma_i'=+1) = \frac{1}{1+e^{-2\beta a_i}} \;, \qquad P(\sigma_i'=-1) = \frac{1}{1+e^{2\beta a_i}} \;. 
 \ee 
In the zero temperature limit we thus have 
 \be
 \begin{split} 
  &a_i>0\,:\qquad \lim_{\beta\rightarrow \infty} P(\sigma_i'=+1) = 1\;, \qquad \lim_{\beta\rightarrow \infty} P(\sigma_i'=-1) = 0\;, \\
  &a_i<0\,:\qquad \lim_{\beta\rightarrow \infty} P(\sigma_i'=+1) = 0\;, \qquad \lim_{\beta\rightarrow \infty} P(\sigma_i'=-1) = 1\;, 
 \end{split} 
 \ee
and so for $T\rightarrow 0$ we recover the deterministic update rule (\ref{UdateRUle}). For any non-zero $\beta$ or non-zero $T$
there is a non-vanishing probability that if the neural network is in an energy minimum it will leave it again.

We expect that after some time the Boltzmann machine will reach thermodynamic equilibrium, by which we mean that any time later 
the probability of finding the system in a certain state or configuration ${\sigma}$ is given by the Boltzmann distribution 
 \be\label{BoltzmannOnceMOre} 
  P_B({\sigma}) = \frac{1}{Z} \, e^{-\beta E({\sigma})}\;, 
 \ee
where we expect the energy function to be given by the spin-glass energy (\ref{SpinGlassBoltzmann}).  
We will not prove this rigorously, but make the following non-trivial consistency check: 
Imagining  that the update rule keeps running after equilibrium 
has been reached, 
we compute the conditional probability 
for neuron $i$ to be in state $\sigma_i=+1$ or $\sigma_i=-1$, \textit{given} that all other neurons are in some prescribed state, and 
we show that this agrees precisely with the above update rule. 

This conditional probability is given by, c.f.~(\ref{TRUEConditionalPROB}), 
 \be\label{condProb} 
 \begin{split} 
  P_c(\sigma_i\big|\sigma_1,\ldots,\sigma_{i-1},\sigma_{i+1},\ldots,\sigma_N) &:=\frac{P_B({\sigma})}
  {\sum_{s=\pm 1}P_B(\sigma_1,\ldots,\sigma_{i-1},s, \sigma_{i+1},\ldots,\sigma_N)} \\
  &= \frac{e^{-\beta E({\sigma})}}{e^{-\beta E(s=+1)} +e^{-\beta E(s=-1)}}\;, 
 \end{split} 
 \ee
where we used that the normalization factor in (\ref{BoltzmannOnceMOre}) cancels, and we introduced the short-hand  
$E(s=\pm 1)$ for the energy where $\sigma_i=s$ and all other $\sigma$ are  given by the prescribed ones with respect to which 
we condition. Precisely as in (\ref{energyDifference2}) we have 
 \be
  E(s=+ 1) - E(s=- 1) = -2\Big(\sum_j w_{ij} \sigma_j + b_i\Big) =-2a_i \;, 
 \ee
where we recalled (\ref{aiiiii22}). We can thus write 
 \be\label{STEPCsdv} 
 \begin{split} 
  e^{-\beta E(s=+1)} +e^{-\beta E(s=-1)} & =  e^{-\beta E(s=+1)}\big(1+e^{-2\beta a_i }\big)  \\
  &= e^{-\beta E(s=-1)}\big(1+e^{2\beta a_i }\big)\;. 
 \end{split} 
 \ee
Returning to (\ref{condProb}) we can use the first line of (\ref{STEPCsdv}) to 
compute the conditional probability that $\sigma_i=+1$:  
 \be
  P_c(\sigma_i=+1 \big|\sigma_1,\ldots,\sigma_{i-1},\sigma_{i+1},\ldots,\sigma_N) 
  =\frac{e^{-\beta E(s=+1)}}{ e^{-\beta E(s=+1)}\big(1+e^{-2\beta a_i }\big) } 
  =\frac{1}{1+e^{-2\beta a_i }}\;. 
 \ee
Similarly, we can use the  the second   line of (\ref{STEPCsdv})  to compute the  conditional probability that $\sigma_i=-1$: 
  \be
  P_c(\sigma_i=-1 \big|\sigma_1,\ldots,\sigma_{i-1},\sigma_{i+1},\ldots,\sigma_N) 
  =\frac{e^{-\beta E(s=-1)}}{ e^{-\beta E(s=-1)}\big(1+e^{2\beta a_i }\big) } 
  =\frac{1}{1+e^{2\beta a_i }}\;. 
 \ee
These agree precisely with the stochastic update rules (\ref{stochupdaterules}).

\bigskip

\noindent \textit{Learning: }\\
After exploring the properties of Boltzmann machines with given weights $w_{ij}$ and bias $b_i$ we turn to the \textit{inverse} 
problem: Given a probability distribution $P_{\rm data}$ (say obtained from a {data} set), how can we \textit{learn} the $w_{ij}$ and $b_i$ 
so that the Boltzmann machine, after reaching  thermodynamic equilibrium, describes approximately this 
 probability distribution? 

We want to minimize some notion  of distance between the Boltzmann distribution (\ref{BoltzmannOnceMOre}) and $P_{\rm data}$. 
A good  measure is the \textit{relative entropy}  or \textit{Kullback-Leibler divergence} defined by  
 \be
  D(P_{\rm data}| P_B) := \sum_{\sigma} P_{\rm data}(\sigma)  \ln \Big(\frac{P_{\rm data}(\sigma)}{P_B(\sigma) }\Big) \geq 0\;. 
 \ee
This equals zero precisely if and only if $P_{\rm data} = P_B$ and is otherwise positive (see the appendix).  
Thus, it is a good strategy to try to minimize this quantity with respect 
to $w_{ij}$ and bias $b_i$. 

To this end we use a notational trick: by extending the index labelling the neurons by one dummy value $0$, $i\rightarrow (i,0)$, 
defining new weights $w_{i0}=w_{0i}=b_i$, and finally fixing the new dummy neuron to $\sigma_0=+1$, we can write the complete 
spin-glass energy as 
 \be
  E(\sigma) = -\frac{1}{2} w_{ij}\sigma_i\sigma_j\;, 
 \ee 
with  the magnetic field term now absorbed into the first term. (For the sake of this computation we also employ the Einstein summation 
convention, in which the sum over $i, j$ is not displayed.)  

In order to minimize  $D\equiv D(P_{\rm data}| P_B)$ with respect to the weights $w_{ij}$ we compute 
 \be\label{partialDD} 
  \frac{\partial D} {\partial w_{ij}} = -\frac{\partial}{\partial w_{ij}} \sum_{\sigma} P_{\rm data} (\sigma) \ln P_B(\sigma)\
  = - \sum_{\sigma} \frac{P_{\rm data}(\sigma)}{P_B(\sigma)}\frac{\partial P_B(\sigma) }{\partial w_{ij}}
  \;, 
 \ee
where we used that only $P_B$ depends on $w_{ij}$. For the partial derivatives one computes 
 \be
 \begin{split} 
  \frac{\partial P_B(\sigma) }{\partial w_{ij}}&= \frac{\partial}{\partial w_{ij}}\Big(\frac{e^{-\beta E(\sigma) }}
  {Z }\Big)
  = \frac{\partial}{\partial w_{ij}}\Big(\frac{e^{\frac{1}{2}\beta w_{kl} \sigma_k\sigma_l}}
  {\sum_{\sigma'}e^{\frac{1}{2}\beta w_{pq} \sigma'_p\sigma'_q} }\Big)\\
  &=\frac{\frac{1}{2}\beta \sigma_i\sigma_j \, e^{\frac{1}{2}\beta w_{kl} \sigma_k\sigma_l}}{Z} 
  -\frac{e^{\frac{1}{2}\beta w_{kl} \sigma_k\sigma_l}}{Z^2} \, 
  \frac{\partial}{\partial w_{ij}}\sum_{\sigma'}e^{\frac{1}{2}\beta w_{pq} \sigma'_p\sigma'_q} \\
  &= \frac{1}{2} \beta \sigma_i\sigma_j P_{B}(\sigma) 
  -P_B(\sigma) 
  \sum_{\sigma'} \frac{1}{2} \beta \sigma_i'\sigma_j' P_{B}(\sigma') \\
  &=\frac{1}{2}\beta P_B(\sigma) \Big(  \sigma_i\sigma_j  - \langle \sigma_i \sigma_j\rangle_{B} \Big) \;, 
 \end{split} 
 \ee
where by $\langle \;\; \rangle_{B}$ we denote the expectation value in the Boltzmann distribution.  
 Inserting this back into (\ref{partialDD}) we have 
  \be
  \begin{split} 
   \frac{\partial D} {\partial w_{ij}} &= -\frac{1}{2}\beta  \sum_{\sigma}
   P_{\rm data}(\sigma)  \Big(  \sigma_i\sigma_j  - \langle \sigma_i \sigma_j\rangle_{B} \Big)\\
   &= -\frac{1}{2}\beta\Big(\langle \sigma_i\sigma_j\rangle_{\rm data}  - \langle \sigma_i \sigma_j\rangle_{B} \Big)\;, 
  \end{split} 
  \ee
where by $\langle \;\; \rangle_{\rm data}$ we denote  the expectation value in the probability distribution $P_{\rm data}$. 

Thus, if we update the weights by gradient descent, i.e., if we update by the rule 
 \be\label{learningBM} 
  \Delta w_{ij} = \epsilon \Big(\langle \sigma_i\sigma_j\rangle_{\rm data}  - \langle \sigma_i \sigma_j\rangle_{B} \Big)\;, 
 \ee
where $\epsilon$ is a learning rate (into which a $\beta$ dependent factor was absorbed), 
we can expect that the probability distribution of the Boltzmann machine gets closer and closer to the given one based on a data set.

\subsection{Restricted Boltzmann Machines: Integrating out Hidden Neurons}

The above learning algorithm is elegant but computationally expensive. This can be improved by turning to 
\textit{restricted Boltzmann machines} (RBM), in which additional so-called \textit{hidden} neurons are added 
that are not visible on the probability space of the prescribed distribution. Moreover, these hidden neurons are not connected among each 
other but only to the remaining visible ones. Due to the reduction of the number of weights this leads to a computational speedup. 
Adding many  layers of  such hidden neurons one obtains \textit{deep Boltzmann machines}, which inspired 
the \textit{deep learning} 
of  modern AI  that we will turn to in the next section. 

To define RBMs we start from the spin/neuron variables $\sigma_I$, which we temporarily label by $I=1,\ldots, N$. 
These are split into two sets, the \textit{visible} ones and the \textit{hidden} ones: 
 \be
  \sigma_I = ( v_i, h_{\alpha})  \;, \qquad i = 1,\ldots, n\;, \quad \alpha=1,\ldots d\;, 
 \ee
where $N=n+d$. So far this is just an arbitrary grouping of the neurons into two subsets, but as alluded to above we will now 
consider the special case of Boltzmann machines in which there are no couplings or weights among the visible or among the hidden neurons, 
but only couplings between them. Thus, the energy function of RBMs takes the form 
 \be\label{RBMEnergy} 
  E(v,h) = -\sum_{i,\alpha} w_{i\alpha} v_i h_{\alpha} - \sum_i b_i v_i -\sum_{\alpha} c_{\alpha} h_{\alpha}\;. 
 \ee
 
Applying the learning rule  (\ref{learningBM}) we could update the weights by 
 \be\label{RBMlearning} 
  \Delta w_{i\alpha} = \epsilon \big(\langle v_i h_{\alpha}\rangle_{\rm data}  - \langle v_i h_{\alpha}\rangle_B\big) \;. 
 \ee
This would just be a special Boltzmann machine with the  a priori assumption that the neurons can be 
grouped into two sets so that there are no couplings between them. However, this idea becomes powerful if we assume that the $h_{\alpha}$ 
are \textit{not} part of the given probability distribution (encoded in data) that we want to model, i.e., $P_{\rm data}(v)$ only 
depends on the visible ones, and the hidden ones play an auxiliary role. 
In the language of RG from the previous section, we have to `integrate out' the hidden neurons 
to obtain the desired probability distribution on the space of $v$'s. Or, to put it differently, given the problem of modeling 
$P_{\rm data}(v)$ it turns out to be greatly beneficial to `integrate in' hidden neurons. 

Concretely, we have to address the following problem: Since  $P_{\rm data}$ is only a function of $v$ the first term 
$\langle v_i h_{\alpha}\rangle_{\rm data}$ in  (\ref{RBMlearning}) is not well-defined. We solve it by determining  the hidden 
variable $h_{\alpha}$ by computing its expectation value with respect to  the \textit{conditional} probability $P(h_{\alpha}|v)$: 
 \be\label{expectationhalpha} 
  \langle h_{\alpha} \rangle_{P(h_{\alpha}|v)} = \sum_{h_{\alpha}=\pm 1} h_{\alpha}  P(h_{\alpha} |v)
  =P(h_{\alpha}=+1|v) - P(h_{\alpha}=-1|v)\;.  
 \ee
It is a nontrivial statement, and only true for the special class of Boltzmann machines, that this conditional probability 
for a fixed $h_{\alpha}$ depends only the $v$'s, not on the remaining hidden neurons.  
 To see this we write out 
 the conditional probability for the entire configuration $h=(h_{\alpha})$ of hidden neurons, which is given by 
  \be\label{condpropsteppp} 
   P(h|v) = \frac{P(v,h)}{\sum_{h'=\pm 1}P(v, h')} = \frac{e^{-\beta E(v,h)}}{\sum_{h'=\pm 1} e^{-\beta E(v,h')}}\;,  
  \ee
 where we used that the normalization factors $\frac{1}{Z}$ cancel. 
In order to compute this we write out with (\ref{RBMEnergy}) 
 \be
 \begin{split} 
  e^{-\beta E(v,h)} &= \exp\big(\beta\sum_i b_i v_i \big) \exp\Big(\beta\sum_{\alpha}\big(\sum_i w_{i\alpha}v_i +c_{\alpha}\big)h_{\alpha}\Big) \\
  &= \exp\big(\beta\sum_i b_i v_i \big) \exp\Big(\beta\sum_{\alpha}a_{\alpha}(v) h_{\alpha} \Big) \;, 
 \end{split} 
 \ee
 where we defined 
  \be\label{aofv} 
   a_{\alpha}(v) := \sum_i w_{i\alpha}v_i +c_{\alpha}\;. 
  \ee 
Writing out the sum in the denominator of (\ref{condpropsteppp}) with this we have 
 \be\label{hsumcondProbandshit} 
 \begin{split} 
  \sum_{\{h_{\alpha}'=\pm 1\}} e^{-\beta E(v,h')} &=  \exp\big(\beta\sum_i b_i v_i \big) \Big(\sum_{h_1'=\pm 1}e^{\beta a_1(v) h_1'}\Big)
  \cdots \Big(\sum_{h_d'=\pm 1}e^{\beta a_d(v) h_d'}\Big)\\
  &=  \exp\big(\beta\sum_i b_i v_i \big) \prod_{\alpha=1}^{d} 2\cosh\big(\beta a_{\alpha}(v)\big) \;, 
 \end{split} 
 \ee
and thus back in (\ref{condpropsteppp}) 
 \be
  P(h|v) := \frac{e^{\beta \sum_{\alpha}a_{\alpha}(v) h_{\alpha}}}{\prod_{\alpha=1}^{d} 2\cosh\big(\beta a_{\alpha}(v)\big) }
  = \prod_{\alpha=1}^{d} P(h_{\alpha}|v) \;, 
 \ee
where we defined 
 \be
 P(h_{\alpha}|v) =  \frac{e^{\beta a_{\alpha}(v) h_{\alpha}}}{2\cosh\big(\beta a_{\alpha}(v)\big) }\;. 
 \ee
This can be interpreted as the conditional probability for a fixed $h_{\alpha}$ \textit{given} that the visible neurons 
are in some state $v$. Finally, we can compute (\ref{expectationhalpha}): 
 \be
 \begin{split} 
  \langle h_{\alpha} \rangle_{P(h_{\alpha}|v)} &=  \frac{e^{\beta a_{\alpha}(v) }}{2\cosh\big(\beta a_{\alpha}(v)\big) }
  - \frac{e^{-\beta a_{\alpha}(v)}}{2\cosh\big(\beta a_{\alpha}(v)\big) }
  =\tanh\big(\beta a_{\alpha}(v)\big)\\
   &= \tanh\Big(\beta \big(\sum_i w_{i\alpha}v_i +c_{\alpha}\big)\Big)\;, 
 \end{split}  
 \ee
where we have recalled (\ref{aofv}). This expression allows us to eliminate $h_{\alpha}$ in terms of a function of the $v_{i}$, 
and using this as 
 \be
  \langle v_i h_{\alpha}\rangle_{\rm data} \ \rightarrow \ \big\langle v_i  \tanh\big(\beta a_{\alpha}(v)\big)\big\rangle_{\rm data} 
 \ee
we can make the update rule (\ref{RBMlearning}) well-defined. 

\bigskip

\noindent \textit{Relation to Renormalization Group: }\\
Above we have used the language of `integrating out' degrees of freedom 
used in the renormalization group (RG). Let us spell out the detailed 
relation \cite{Mehta:2014xqf,Iso:2018yqu,Koch:2019fxy}. 

The sum defining the  partition function based on (\ref{RBMEnergy}), 
 \be
  Z = \sum_{v,h} e^{-\beta E(v,h)} 
 \ee
can be partially  performed by summing over the hidden variables $h$: 
 \be\label{partialhSum} 
  e^{d g }  e^{-\beta \widetilde{E}(v)} := \sum_{h} e^{-\beta E(v,h)} \;, 
 \ee
where in parallel to the RG treatment of the Ising model in (\ref{RGedIsingSum}) $e^{d g } $ is defined implicitly by not depending on the visible neurons $v$. 
The exact partition function is then given by 
 \be\label{exactPartition} 
  Z = \sum_{v} e^{-\beta \widetilde{E}(v) +dg} \;. 
 \ee
This is exactly the RG procedure, except that the integrating out of the $h_{\alpha}$ is not 
directly or a priori related to a change of lattice spacing. 

Now, the sum (\ref{partialhSum}) is precisely the one we have already performed in (\ref{hsumcondProbandshit}):
 \be
 \begin{split} 
  \sum_{h} e^{-\beta E(v,h)} &= \exp\Big( \beta\sum_i b_i v_i + \ln  \prod_{\alpha} 2\cosh(\beta a_{\alpha}(v))\Big) \\
  &=  \exp\Big( \beta\sum_i b_i v_i +   \sum_{\alpha} \ln \big(2\cosh(\beta a_{\alpha}(v))\big) \Big) \\
  &= e^{\ln(2) d} \exp\Big( \beta\sum_i b_i v_i +   \sum_{\alpha} \ln \big(\cosh(\beta a_{\alpha}(v))\big) \Big)\;, 
 \end{split} 
 \ee
from which we read off by comparison with (\ref{partialhSum}) that $g=\ln 2 $ and 
 \be\label{exactintegratedoutenergy} 
  \widetilde{E}(v)  = - \sum_i b_i v_i -\frac{1}{\beta}  \sum_{\alpha} \ln \big(\cosh(\beta a_{\alpha}(v))\big) \;. 
 \ee 

For the special case of restricted Boltzmann machines we were able to integrate out the hidden neurons exactly, 
but the  resulting expression (\ref{exactintegratedoutenergy}) is unfamiliar due to the presence of logarithmic and hyperbolic functions. 
We can, however, gain some intuition by Taylor expanding, assuming small $\beta$, 
which to first non-trivial order yields (c.f.~(\ref{lncoshexpansion})): 
 \be\label{exactintegratedoutenergyexpandedsdcsd} 
 \begin{split} 
  \widetilde{E}(v)  &= - \sum_i b_i v_i -\frac{1}{2\beta}  \sum_{\alpha}(\beta a_{\alpha}(v))^2+ \cdots \\
  &= - \sum_i b_i v_i -\frac{1}{2}  \sum_{\alpha}\beta \big(\sum_i w_{i\alpha}v_i +c_{\alpha}\big) ^2+ \cdots
   \;. 
 \end{split} 
 \ee 
Thus, to leading order, the integrating out of hidden neurons $h_{\alpha}$ induces  the energy function 
 \be
  \widetilde{E}(v) = -\frac{\beta}{2} \sum_{\alpha} c_{\alpha}^2-\frac{1}{2} \sum_{i,j} \widetilde{w}_{ij} v_i v_j - \sum_{i}\widetilde{b}_i v_i 
  +\cdots \;,  
 \ee 
where the first term is an irrelevant constant term, and 
 \be
  \widetilde{w}_{ij} := \beta \sum_{\alpha} w_{i\alpha} w_{j\alpha} \;, \qquad 
  \widetilde{b}_{i} := b_i +\beta \sum_{\alpha} w_{i\alpha} c_{\alpha}\;. 
 \ee
We obtain a spin-glass type model for the visible neurons $v_i$, with the weight matrix obtained from the 
original couplings between visible and hidden neurons by the above formula and with an `effective' external magnetic field 
containing the original one coupling to the hidden neurons.

Finally note that Taylor expanding (\ref{exactintegratedoutenergy}) 
to yet higher order in $\beta$ would induce higher-order terms in $v_i$ of the structural form 
 \be
  \widetilde{E}(v) = -\frac{1}{2} \sum_{i,j} \widetilde{w}_{ij} v_i v_j - \sum_{i}\widetilde{b}_i v_i
  + \frac{1}{3!} \sum_{i,j,k} A_{ijk} v_i v_j v_k + \sum_{i,j,k,l} B_{ijkl} v_i v_j v_k v_l+\cdots   \;,  
 \ee 
We could have started right away with the model of this general form but then the problem would present itself how to 
learn all the weights $w_{ij}$, $A_{ijk}$, $B_{ijkl}$, etc., which presumably would require learning rules such as (\ref{learningBM}), 
in turn needing one to compute  3-point, 4-point and higher order correlation functions. Clearly, this becomes 
computationally intractable. The power of restricted Boltzmann machines lies in the basic energy function being quadratic 
in the (visible and hidden) spin variables, with the higher-order weight matrices only arising upon integrating out the hidden neurons. 
While the modern `deep learning' techniques at the heart of the AI revolution use little detailed techniques of Boltzmann machines 
this basic idea remains: to encode hidden regularities and correlations in layers of hidden neurons that do not participate 
directly in the data or function being learned.

\section{Deep Learning and Large Language Models}

In this section we explain  the basic mathematical techniques underlying the \textit{deep learning}  based 
on a large number of hidden neurons, and we introduce the special class of large language models that are the basis of 
the current  AI developments.

\subsection{Deep Learning}

\begin{figure}
    \centering
    \includegraphics[width=0.35\textwidth]{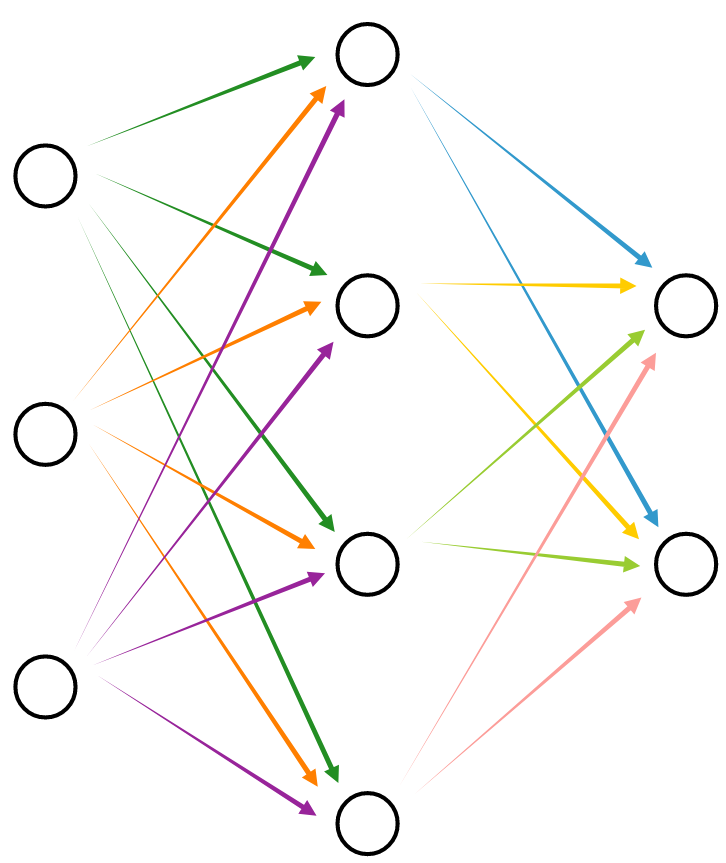}
    \caption{A feedforward network with one hidden layer, here representing  a 
    function $\mathbb{R}^3\rightarrow \mathbb{R}^4\rightarrow \mathbb{R}^2$ }
    \label{Deep}
\end{figure}

The basic deep learning architecture is a \textit{feedforward network} consisting of several layers of neurons, as depicted in figure \ref{Deep} 
for one hidden layer (although in practice deep learning uses more than one hidden layer, see \cite{Nielsen} for an excellent introduction to deep learning).  
The neurons are only loosely connected to the neurons of the spin-glass type Boltzmann machines:  they 
do not need to take the values $+1$ or $-1$. Rather, in each layer they represent real-valued components of a vector belonging 
to, say, $\mathbb{R}^{k}$ if there are $k$ neurons  in the given layer. 
We label the layers of the network by $\ell=1,2,\ldots $ and denote the vector space corresponding to each layer by $\mathbb{R}^{n_{\ell}}$. 
The first layer may be called the input layer, as it represents 
the input vector for a nonlinear function that the network computes, and the last layer may be called the output layer. 
All other layers are \textit{hidden layers}. 
As before, this neural network is equipped with certain 
weights, which here are matrices and bias vectors associated to each hidden layer and the output layer: 
 \be
  \text{weights:}\qquad W^{(\ell)} = \big(w^{(\ell)}_{ij}\big) \;, \quad 
  {\bf b}^{(\ell)} = \big(b_{i}^{(\ell)}\big)\;, \quad \ell=2,3,\ldots, L\;, 
 \ee
where $L$ is the total numbers of layers. Note that the range of the indices $i,j,\ldots $ varies from layer to layer: 
in general the matrices $W^{(\ell)}$ are not square matrices.

The neurons in each layer $\ell >  1$ receive inputs from all neurons of the 
previous layer $\ell-1$, and the following iterative computation is performed: Starting with the 
\textit{input} vector ${\bf x}$ in the first layer we set 
 \be\label{COmpSTEPs1} 
  {\bf a}^{(1)}  : =  {\bf x}  \in \mathbb{R}^{n_1} \;, 
 \ee
 then compute for the next layers $\ell\geq 2$ 
 \be\label{COmpSTEPs2} 
  z^{(\ell)}_i:= \sum_j w^{(\ell)}_{ij} a^{(\ell-1)}_{j}  +b_i^{(\ell)}\quad \Leftrightarrow \quad 
   {\bf z}^{(\ell)}:= W^{(\ell)} {\bf a}^{(\ell-1)}  + {\bf b}^{(\ell)}\;, 
 \ee 
and act with a nonlinear function $\phi$ component-wise to determine 
 \be\label{COmpSTEPs3} 
   {\bf a}^{(\ell)}:=\phi({\bf z}^{(\ell)}) \quad \Leftrightarrow \quad  a^{(\ell)}_i := \phi(z^{(\ell)}_i) \;. 
 \ee   
The nonlinear function is often taken to be the sigmoid or logistic function: 
 \be
   \phi(z):=\sigma(z)=\frac{1}{1+e^{-z}}\;, 
 \ee
but we will allow generic nonlinear  functions for the following discussion.  
Iterating the above computation by moving through the network  `forward', layer by layer, 
 we finally obtain the output vector of the neural network (the value of 
the function $f$ computed by this network): 
 \be\label{NNFunction} 
  { f}_{\Theta}({\bf x}) := {\bf a}^{(L)} \in \mathbb{R}^{n_L}\;, 
 \ee
where we denoted the weights collectively by $\Theta=(w^{(\ell)}_{ij}, b_i^{(\ell)})$.  
 
 This is it. The remarkable fact is that this simple architecture can perform all kinds of remarkable 
 operations. For instance, the input vector ${\bf x}$ could encode the pixels of a photo and the output could  be 
 $0$ or $1$ to indicate whether the photo contains a cat (yes or no).  
There are universal approximation theorems that guarantee that a network with a single hidden layer can approximate arbitrary 
continuous functions \cite{Cybenko1989,HornikStinchcombeWhite1989}, but this may require an impractically large 
number of hidden neurons. 
Empirically one finds that increasing the depth yields vastly more efficient networks.   
 The crucial question is how to determine the above weights (to train the network) so that it performs a desired operation. 
 This requires training data and the backpropagation algorithm to which we turn next.

\subsection{Backpropagation}

The inverse problem we want to solve is the following: Suppose we are given a data set for the desired function, i.e., 
we are given a (sufficiently large) number $m$ of pairs 
 \be\label{DATAAAAAA} 
  ({\bf x}_{\alpha}, {\bf y}_{\alpha}) \ \in \ \mathbb{R}^{n_1}\times \mathbb{R}^{n_L} \;, \quad \alpha=1,\ldots, m\;,  
 \ee
how can we fix the weights $\Theta$ so that (\ref{NNFunction}) encodes, to a good approximation, 
this data set, i.e., $f_{\Theta}({\bf x}_{\alpha})\simeq {\bf y}_{\alpha}$ for all $\alpha$? 
Once this inverse problem is solved,  
the hope is then 
that this network will accurately compute this function on new examples (say photos with or without cats on 
which the network was not trained).  

To this end we need a \textit{cost} or \textit{loss}  function that measures how much the function computed by the network for a given set 
of weights (say chosen at random) differs from the desired one. 
For instance, a canonical cost function is 
 \be\label{CostFunction} 
  {\cal C}(\Theta):= \frac{1}{2m} \sum_{\alpha=1}^{m} || {\bf y}_{\alpha}-f_{\Theta}({\bf x}_{\alpha})||^2
  = \frac{1}{2m} \sum_{\alpha=1}^{m} || {\bf y}_{\alpha}-{\bf a }^{(L)}_{\alpha}||^2
  = \frac{1}{m} \sum_{\alpha=1}^{m} C_{\alpha}(\Theta) \;, 
 \ee
where in the second equality we used (\ref{NNFunction}), with the ${\bf a }^{(\ell)}_{\alpha}$ being the vectors obtained from 
(\ref{COmpSTEPs1})--(\ref{COmpSTEPs3}) by starting with ${\bf x}_{\alpha}$, 
and in the last equality we defined\footnote{In the literature cost and loss function are sometimes distinguished as follows: 
the loss function is evaluated on a single training example ($C_{\alpha}(\Theta)$ for a fixed $\alpha$), 
while the cost function is the average ${\cal C}(\Theta)$ over the entire training data.} 
 \be
  C_{\alpha}(\Theta) = \frac{1}{2} || {\bf y}_{\alpha}-{\bf a }^{(L)}_{\alpha}||^2\;. 
 \ee
In (\ref{CostFunction}) we displayed the dependence on the weights $\Theta$ since we want to minimize the cost function with 
respect to the weights. Of course, the values of the cost function also depend on the training data but for the sake of this computation 
these are considered fixed. 

In order to minimize  the cost function we will use gradient descent. We need the gradient of the cost function 
with respect to all the weights, i.e.~we need to compute 
 \be
   \frac{\partial C}{\partial w_{ij}^{(\ell)}}  \quad \text{and} \quad  \frac{\partial C}{\partial b_{i}^{(\ell)}}\;,
 \ee   
for all $i, j$ and $\ell$. Here we momentarily suppress the dependence on the training data index $\alpha$. 
As given, $C$ only depends on the ${\bf a}^{(L)}$ and hence ${\bf z}^{(L)}$ of the last layer, 
but we can make massive use of the chain rule to compute the derivative with respect to all weights. 
For ease of notation  we define the following vectors for each layer: 
 \be\label{Deltaell} 
  {\bf \Delta}^{(\ell)} = (\Delta^{(\ell)}_i) := \frac{\partial C}{\partial z_i^{(\ell)}}\;. 
 \ee 
The vector corresponding to the last or output layer we can immediately compute with the chain rule: 
 \be\label{Vecdfjegeg} 
  \Delta_i^{(L)}:=\sum_j \frac{\partial C}{\partial a_{j}^{(L)}} \frac{\partial a_{j}^{(L)}}{\partial z_{i}^{(L)}} 
  =\sum_j \frac{\partial C}{\partial a_{j}^{(L)}} \delta_{ij} \phi'(z_{i}^{(L)}) = \frac{\partial C}{\partial a_{i}^{(L)}}\phi'(z_{i}^{(L)}) \;, 
 \ee
where we used (\ref{COmpSTEPs3}) (in particular that the dependence of $a$ on $z$ is component-wise, 
so that $\frac{\partial a_{j}^{(\ell)}}{\partial z_{i}^{(\ell)}}$ can only be non-zero for $i=j$, bringing in the Kronecker delta).  
 In order to write this relation in matrix/vector notation we need to define the \textit{Hadamard product} between two vectors 
 ${\bf a}, {\bf b}\in \mathbb{R}^k$, which is just the vector whose components are the product of the components: 
  \be
   ({\bf a} \odot {\bf b})_i = a_i b_i\qquad \text{(no sum)}\;. 
  \ee
Then (\ref{Vecdfjegeg}) can be written as 
 \be\label{LlayerGrad} 
  {\bf \Delta}^{(L)} = {\rm grad}_{{\bf a}^{(L)}}C \odot \phi'({\bf z}^{(L)})\;. 
 \ee
 
Next we can  compute the vectors (\ref{Deltaell}) 
for all layers. Using the chain rule again we can determine $\Delta^{(\ell)}$ from $\Delta^{(\ell+1)}$ for any $\ell$: 
\be\label{Deltajelllll} 
 \Delta_i^{(\ell)} = \sum_j \frac{\partial C}{\partial z_j^{(\ell+1)}} \frac{\partial z_j^{(\ell +1)}}{\partial z_i^{(\ell)}}
 = \sum_j \Delta_j^{(\ell+1)} \frac{\partial z_j^{(\ell +1)}}{\partial z_i^{(\ell)}}\;. 
\ee
In order to compute the remaining  partial derivative we recall with (\ref{COmpSTEPs2}) and (\ref{COmpSTEPs3}) that 
 \be
  z_j^{(\ell+1)} = \sum_k w_{jk}^{(\ell+1)} \phi(z_k^{(\ell)}) +b_j^{(\ell+1)} \;, 
 \ee
so that 
 \be
   \frac{\partial z_j^{(\ell +1)}}{\partial z_i^{(\ell)}}  = \sum_{k} w_{jk}^{(\ell+1)}\phi'(z_k^{(\ell)})\frac{\partial z_k^{(\ell)}}{\partial z_i^{(\ell)}} 
   = w_{ji}^{(\ell+1)}\phi'(z_i^{(\ell)})\;. 
 \ee
Reinserting this into (\ref{Deltajelllll}) we have 
 \be
   \Delta_i^{(\ell)} =  \sum_j w_{ji}^{(\ell+1)} \Delta_j^{(\ell+1)} \phi'(z_i^{(\ell)})\;, 
 \ee
or in matrix/vector notation: 
 \be\label{Deltaellgeneral} 
  {\bf \Delta}^{(\ell)} = \big((W^{(\ell+1)})^T{\bf \Delta}^{(\ell+1)}\big)\odot \phi'({\bf z}^{(\ell)})\;. 
 \ee
 
Finally, we can use the chain rule compute the gradient of the cost function $C$ with respect to the weights 
$w$ and $b$: 
 \be\label{FINALGrad} 
 \begin{split} 
  \frac{\partial C}{\partial w_{ij}^{(\ell)}} &= \sum_k \frac{\partial C}{\partial z_k^{(\ell)}} \frac{\partial z_k^{(\ell)}}{\partial w_{ij}^{(\ell)}}
  = \frac{\partial C}{\partial z_i^{(\ell)}} a_j^{(\ell-1)}
  = \Delta_i^{(\ell)}  a_j^{(\ell-1)} \;, \\
   \frac{\partial C}{\partial b_{i}^{(\ell)}} &= \sum_j \frac{\partial C}{\partial z_j^{(\ell)}}\frac{\partial z_j^{(\ell)}}{\partial b_i^{(\ell)}} 
   = \frac{\partial C}{\partial z_i^{(\ell)}} = \Delta_i^{(\ell)} \;. 
 \end{split}  
 \ee 
where we used (\ref{COmpSTEPs2}). This solves our problem of finding the gradient of $C$ with respect to the weights. 

\medskip

Let us  summarize the training process using the \underline{backpropagation algorithm} \cite{HintonBackProp}: 
\begin{itemize}

\item[1)] \textit{Input:} 
 \be
   {\bf x} =: {\bf a}^{(1)} \;\;(\ell=1)
 \ee  
\item[2)] \textit{Feedforward:} For $\ell=2,3,\ldots L$ compute 
 \be
  {\bf a}^{(\ell)} = \phi({\bf z}^{(\ell)})\;, \quad {\bf z}^{(\ell)} = W^{(\ell)} {\bf a}^{(\ell-1)} +{\bf b}^{(\ell)} \;. 
 \ee
\item[3)] \textit{Output Error:} Compute (\ref{LlayerGrad}), 
 \be\label{LlayerGrad334} 
  {\bf \Delta}^{(L)} = {\rm grad}_{{\bf a}^{(L)}}C \odot \phi'({\bf z}^{(L)})\;.   
 \ee
 
 \item[4)] \textit{Backpropagate Error:} For $\ell=L-1, L-2,\ldots, 2$ compute ${\bf \Delta}^{(\ell)}$ with (\ref{Deltaellgeneral}), 
 \be\label{Deltaellgeneral453w45} 
  {\bf \Delta}^{(\ell)} = \big((W^{(\ell+1)})^T{\bf \Delta}^{(\ell+1)}\big)\odot \phi'({\bf z}^{(\ell)})\;. 
  \ee
  
  \item[5)] \textit{Output Gradients:} Compute the gradients with (\ref{FINALGrad}), 
   \be\label{FINALGRADIENTSSS} 
    \frac{\partial C}{\partial w_{ij}^{(\ell)}}   = \Delta_i^{(\ell)}  a_j^{(\ell-1)} \;, \qquad 
   \frac{\partial C}{\partial b_{i}^{(\ell)}} = \Delta_i^{(\ell)} \;. 
  \ee
 
\end{itemize}

Note that in order to compute the final gradients by backpropagating the error 
one needs the  ${\bf a}^{(\ell)}$ and ${\bf z}^{(\ell)}$ that were 
computed in the feedforward part and that hence need to be saved.

In the above discussion we have not indicated explicitly the index $\alpha$ and hence which data point is 
taken as the input vector and with respect to which output vector we compute the output error. 
In practice one computes the gradient (\ref{FINALGRADIENTSSS}) for each of the $m$ data points (\ref{DATAAAAAA}) 
and then updates the weights by 
 \be
  w_{ij}^{(\ell)} \rightarrow  w_{ij}^{(\ell)} -\frac{\epsilon}{m} \sum_{\alpha=1}^{m} \Delta_{\alpha i}^{(\ell)}  a_{\alpha j}^{(\ell-1)}\;, \qquad
  b_{i}^{(\ell)} \rightarrow  b_{i}^{(\ell)}-\frac{\epsilon}{m} \sum_{\alpha=1}^{m}\Delta_{\alpha i}^{(\ell)}\;, 
 \ee
where $\epsilon$ is the learning rate.  This gives one step of the gradient descent. 
For the next step we use the updated weights and go through the backpropagation algorithm again, using the 
same data set  (\ref{DATAAAAAA}). One iterates this until the weights converge to a fixed point.

\subsection{Large Language Models}

Large language models (LLMs) are special deep learning architectures that are designed for language applications: 
given a sequence of words, the LLM  predicts the next word. 
(See \cite{Douglas:2023olt,Phuong} for introductions to LLMs and \cite{Bahdanau:2014ghw} for earlier work.) 
For instance, given the following string of five words: 
 \be\label{Sentence} 
  \text{The capital of Germany is}
 \ee
the LLM should predict the word ``Berlin". More precisely, the input of an LLM is a sequence of \textit{tokens}, which are 
words, or parts of words, and the output is a probability 
distribution on this space of tokens. For instance, if the input consists of the five tokens in (\ref{Sentence}) the LLM should assign 
the highest probability to ``Berlin", a much smaller probability to ``Bonn" and a yet much smaller probability to ``Paris". 

More formally, we assume  a fixed \textit{vocabulary} of a number $V$ of tokens: 
 \be\label{voctokens} 
  {\cal V} :=\big\{ w_a \,\big| \, a=1,\ldots, V\big\} \;, 
 \ee
where typically $V$ could be of the order $10^5$. As mentioned, these tokens may contain the commonly used words, 
say in English or German, but also parts of words such as ``ing" in English, punctuation signs or even the empty space between words. 
A sentence as 
 \be
  \text{I like cats.}
 \ee
could then be resolved into five tokens as 
 \be
  \big(``{\rm I}",``{\rm like}", ``{\rm cat}",``{\rm s}",``{\rm .}"\big)\;, 
 \ee
it being understood that between two words there is always an empty space.  
For simplicity  we may just assume that the tokens are complete words.  
Of course, internally  each token $w_a$ may just be represented by the number $a$ of its position in 
the vocabulary, so we can identify ${\cal V}\simeq\{1,2,\ldots, V\}$. 

More generally, given the conditional probability for the next token, \textit{given} the previous tokens, that the LLM provides, one 
can compute  the probability for any string of tokens, 
 \be\label{basicporb} 
  P(x_1,x_2,\ldots,x_n) \;, 
 \ee
where all arguments belong to the vocabulary (\ref{voctokens}) of tokens: $x_i\in {\cal V}$ for $i=1,\ldots, n$.  
Here we use letters for token and index that are different from (\ref{voctokens}). 
The reason is that we want to make clear that the $i$ indicates the position 
of the token within the sequence of tokens 
(say the position of a word in a sentence), which of course has nothing to do with the position $a$ of the 
token within the fixed vocabulary. The LLM computes  the conditional probability 
that given $k-1$ tokens the next token is $x_k$: 
 \be
  P(x_k|x_1,x_2, \ldots, x_{k-1})\;, 
 \ee
and in terms of this the probability  (\ref{basicporb}) can be determined 
as follows:
 \be\label{chainrule} 
 \begin{split} 
  P(x_1,x_2,\ldots, x_n) &= \prod_{k=1}^{n} P(x_k|x_1,x_2, \ldots, x_{k-1}) \\
  &= P(x_1)P(x_2|x_1)P(x_3|x_1,x_2)\cdots P(x_n|x_1,\ldots, x_{n-1}) 
  \;. 
 \end{split} 
 \ee
(See (\ref{TRUEConditionalPROB}) in the appendix for a derivation of this so-called chain rule.)

In the following we give a simplified version of how the transformer architecture of an LLM works in detail 
(the main simplification being the focus on what is called single-head attention rather than multi-head attention) \cite{Transformer}:
\\[1ex] 
\textit{Embedding:} There is an embedding (or representation) for each token as a vector in some $N$-dimensional vector space. Thus, each token is mapped to 
 \be\label{EMBEDDING} 
  w_a \mapsto {\bf e}_a\in \mathbb{R}^{N}\;. 
 \ee
We can think of this as a  list $({\bf e}_a)_{a=1,\ldots, V}$ of $V$ vectors in $\mathbb{R}^{N}$, which in turn defines  a 
$V\times N$ matrix: 
${\bf E} \in\mathbb{R}^{V\times N}$. (Representing the token $w_a$ as the canonical basis vector $\widehat{\bf e}_a\in \mathbb{R}^{V}$, which 
has all components equal to zero, except the $a$ component that is equal to one, the embedding (\ref{EMBEDDING}) can be written as 
the matrix operation ${\bf e}_a= {\bf E}\,\widehat{\bf e}_a$. Typically, $N$ is much less than $V$, so we embed the tokens in a space of 
much smaller dimension.)

\noindent  
Given a sequence of an arbitrary number $n$ of tokens, $(x_1, x_2,\ldots, x_n ) = (w_{a_1}, w_{a_2},\ldots,  w_{a_n}) \in {\cal V}$, 
where $a_1, a_2, \ldots, a_n\in\{1,2,,\ldots, V\}$ give their positions within the fixed vocabulary, we then associate the 
corresponding list of embedding vectors \textit{shifted by a so-called positional embedding vector} ${\bf p}_k\in \mathbb{R}^{N}$
(that is necessary to take into account that the meaning of a word typically depends on its position in a sentence): 
 \be\label{embeddedTokens} 
  (x_1, x_2,\ldots, x_n ) \ \mapsto \ 
  ({\bf h}_1,{\bf h}_2, \ldots, {\bf h}_n) := ({\bf e}_{a_1}+{\bf p}_1, {\bf e}_{a_2}+{\bf p}_2,\ldots,  {\bf e}_{a_n}+{\bf p}_n)\;. 
 \ee 
 In practice the positional embedding vectors are typically learned  during training much like the embedding matrix 
 ${\bf E} \in\mathbb{R}^{V\times N}$.\footnote{As nicely illustrated at \textit{https://www.youtube.com/watch?v=wjZofJX0v4M}, the embedding vectors encode the basic semantic meaning of each word in the Euclidean geometry of this embedding  space  in such a way 
 that one has approximate relations like 
  \be
   {\bf e}({\rm Sushi}) + {\bf e}({\rm Germany}) - {\bf e}({\rm Japan}) \simeq {\bf e} ({\rm Bratwurst})\;. 
  \ee  } 
 They can be grouped into a matrix ${\bf P}\in \mathbb{R}^{K_{\rm max}\times N}$, 
 where $K_{\rm max}$ gives a limit to how many such positional embedding vectors are included.

 \medskip

 \noindent \textit{Contextualization:} The list of vectors (\ref{embeddedTokens}) representing the (embedded) list of tokens  
is  the input of the deep neural network, which is called the  transformer. 
As always, it consist of several layers labelled by $\ell=1,\ldots, L$, 
with  $\ell=1$ being the input layer. 
 What is special here is that we have three kind of weight matrices associated to each hidden layer $\ell=2,\ldots, L-1$: 
 $W_Q^{(\ell)}$, $W_K^{(\ell)}$ and $W_V^{(\ell)}$ called \textit{query}, \textit{keys} and \textit{values}, respectively, 
 which as always  need not be square matrices. 
 Now for each vector in (\ref{embeddedTokens})  one computes for the second layer\footnote{Strictly speaking, this describes only so-called 
 \textit{single head attention}. The multi-head attention used in real LLMs will be described below.}
  \be\label{weightmatricesLLM} 
   \begin{split}
    {\bf q}_{k} := W^{(2)}_Q\,{\bf h}_{k}\;, \qquad {\bf k}_k:= W^{(2)}_K\,{\bf h}_{k}\;, \qquad  {\bf v}_k:= W^{(2)}_V\,{\bf h}_{k}\;,
   \end{split}
  \ee 
and then sets 
 \be\label{hhoch2} 
  {\bf h}^{(2)}_{k}:=\sum_{l=1}^{k}\frac{\exp\big(\frac{1}{\sqrt{d_k}}\langle {\bf q}_k,{\bf k}_l\rangle\big)}
  {\sum_{m=1}^k\exp\big(\frac{1}{\sqrt{d_k}} \langle {\bf q}_k,{\bf k}_m\rangle\big)}\, {\bf v}_l\;, 
 \ee 
where $d_k:={\rm dim}({\bf q}_{k})$ denotes the dimension of the vector space to which ${\bf q}_k$ 
belongs.\footnote{Note that here we impose $l\leq k$, which is a so-called causal mask.}

\medskip

\noindent \textit{Multi-head attention:} 
The previous step described single-head attention, but as an aside, and for completeness, 
we now briefly describe 
the multi-head attention of real LLMs. Multi-head attention
uses several copies of weight matrices, labelled by 
$h=1,\ldots,H$, where $H$ is the number of so-called heads:
\be
W^{(\ell),h}_Q\;,\qquad W^{(\ell),h}_K\;,\qquad W^{(\ell),h}_V\;, 
\qquad h=1,\ldots,H\;. 
\ee
Each is applied as in (\ref{weightmatricesLLM}) to the vectors ${\bf h}_k$ to produce 
\be
{\bf q}_{k}^{h}:= W^{(\ell),h}_Q\,{\bf h}_{k}\;,\qquad 
{\bf k}_{k}^{h}:= W^{(\ell),h}_K\,{\bf h}_{k}\;,\qquad 
{\bf v}_{k}^{h}:= W^{(\ell),h}_V\,{\bf h}_{k}\;. 
\ee
Next, (\ref{hhoch2}) is applied for each $h=1,\ldots,H$, i.e., head by head, to produce 
an  output vector, which we call  $\widetilde{\bf h}_{k}^{h}$.
We assume that the output dimension $d_V$ of the $W_V$ is chosen such 
that $d_V=N/H$, so that the $H$ outputs can be concatenated 
into a single $N$-dimensional vector
\be
\widetilde{\bf h}_{k}:=\big(\widetilde{\bf h}_{k}^{1},\widetilde{\bf h}_{k}^{2},
\ldots,\widetilde{\bf h}_{k}^{H}\big)\,\in\,\mathbb{R}^{N}\;.
\ee
Finally, a linear transformation with a learned $N\times N$ matrix 
$W^{(\ell)}_O$ is applied: 
\be
{\bf h}^{(\ell)}_{k}:= W^{(\ell)}_O\,\widetilde{\bf h}_{k}\;, 
\ee
which replaces (\ref{hhoch2}). 
The motivation for multi-head attention is that different heads can 
attend to different aspects (syntactic or semantic) of the context.  
The multi-head case reduces to the single-head case for $H=1$ 
as then  $W^{(\ell)}_O$ can 
be absorbed into $W^{(\ell)}_V$.

  \medskip

 \noindent \textit{Feedforward through the layers:} 
 Returning to the single-head case, with (\ref{hhoch2}) we  produced a new list of vectors $({\bf h}^{(2)}_1, \ldots, {\bf h}^{(2)}_n)$, and  
taking this as input for the third layer (instead of (\ref{embeddedTokens})) and performing the same operation we move through the network as before 
until the final layer, in each layer producing a list of vectors  
 \be\label{contextualizedlist} 
  ({\bf h}^{(\ell)}_1, {\bf h}^{(\ell)}_2, \ldots, {\bf h}^{(\ell)}_n)\;, \qquad \ell=1,\ldots, L\;, 
 \ee
where $\ell=1$ denotes the initial input list (\ref{embeddedTokens}). 
  \medskip

 \noindent \textit{Output:} Associated to the last layer we have a matrix $W_{\rm out}$ and a bias vector ${\bf b}_{\rm out}$, 
 and taking the last entry of (\ref{contextualizedlist}) for $\ell=L$ one then computes 
  \be\label{lastlayerz} 
   {\bf z}:= W_{\rm out} {\bf h}_n^{(L)} + {\bf b}_{\rm out} \ \in \ \mathbb{R}^V \;, 
  \ee
where we assume the dimensions of $W_{\rm out}$ and  ${\bf b}_{\rm out}$ are such that ${\bf z}\in \mathbb{R}^{V} $. 

The conditional probability that the next token is $w_a$ for a fixed $a$, \textit{given} that the first $n$ tokens are 
given by $(x_1, x_2,\ldots, x_n ) = (w_{a_1}, w_{a_2},\ldots,  w_{a_n})$,  is then 
  \be\label{FinalLLMProb} 
   P(w_a|w_{a_1}, w_{a_2},\ldots,  w_{a_n}) = {\rm softmax}({\bf z})_a:= \frac{e^{z_a}}{\sum_{b=1}^{V} e^{z_b}} \;. 
  \ee
 The softmax function defined above transforms a list of numbers,  here the components $z_a$ of the vector ${\bf z}$, 
 so that they become a probability distribution, i.e., a list of positive numbers that sum to one.   
  
 \medskip

 This completes our (simplified) description of how a LLM operates once the weights are fixed. 
 These weights include the components of the matrices and vectors in (\ref{weightmatricesLLM}) and (\ref{lastlayerz}) 
 but also the embedding matrix ${\bf E} \in\mathbb{R}^{V\times N}$ and 
 the positional embedding vectors ${\bf P}\in \mathbb{R}^{K_{\rm max}\times N}$. 
 These have to be learned from training data, say a large collection (corpus) of text. 
 If the corpus consists of sequences of tokens 
 \be 
   x^{(a)}=\big(x^{(a)}_1,\dots,x^{(a)}_{n_a}\big)\;, \qquad  a=1,\dots,M\;, 
 \ee  
one may define the loss function given by the so-called \textit{cross entropy}:\footnote{See appendix \ref{Shannon} 
for the general definition of cross entropy. To compare with (\ref{DEFCross}) fix the corpus label $a$ and the position $k$ in the string 
of tokens and define $Q_{a,k}:=P_\Theta\!(\;\cdot\,  \mid x^{(a)}_1,\dots,x^{(a)}_{k-1})$. Furthermore, define 
in terms of the Kronecker delta 
 $P_{a,k}(w):=\delta_{w,x_k^{(a)}}$, which  is thus equal to $1$ if $w$ actually equals the next token $x_k^{(a)}$ of the corpus and zero otherwise. 
The cross entropy (\ref{DEFCross}) then reduces to 
 \be
  S(P,Q) = -\sum_w P_{a,k} (w) \ln Q_{a,k}(w) = -\ln P_\Theta\!\big(x^{(a)}_k \mid x^{(a)}_1,\dots,x^{(a)}_{k-1}\big)\,. 
 \ee
 In (\ref{LLMLOSSSSSS}) this is summed over all positions and averaged over the corpus.}
\be\label{LLMLOSSSSSS}
{\cal L}
=
-\frac{1}{M}\sum_{a=1}^M \sum_{k=1}^{n_a}
\ln P_\Theta\!\big(x^{(a)}_k \mid x^{(a)}_1,\dots,x^{(a)}_{k-1}\big)\;, 
\ee
where we made the dependence on the parameters, collectively denoted by $\Theta$, manifest. 
 Using a version of backpropagation one can use gradient descent to adjust the weights and 
 to minimize this cross entropy.

 \medskip
 
 We close by circling back to statistical physics and the Boltzmann-Gibbs distribution, which is not just of academic interest for the 
 following reason: Sticking to the probability distribution  (\ref{FinalLLMProb}), i.e., picking the next word strictly in 
 the frequency predicted by this distribution,  is too rigid  in practice (it produces boring text). 
 Rather, it is advantageous to introduce a temperature parameter 
 that  softens the distribution. Defining the energy function on single tokens as 
  \be
   E(w_a|w_{a_1}, w_{a_2},\ldots,  w_{a_n}) :=  -z_a = -(W_{\rm out} {\bf h}_n^{(L)})_a - ({\bf b}_{\rm out})_a\;, 
  \ee 
in which we should think of the initial tokens as parameters, 
the above probability distribution takes the form of  the Boltzmann distribution: 
 \be
  P(w_a)  = \frac{1}{Z} e^{-\beta E(w_a)}\;, \qquad Z= \sum_{b}e^{-\beta E(w_b)}\;,  
 \ee
with the inverse temperature parameter $\beta$. Setting the inverse  temperature to $\beta=1$  one recovers the original distribution 
(\ref{FinalLLMProb}), but turning up the temperature a bit (lowering $\beta$ a bit) leads to a more useful model. 

As emphasized, in the above we take the initial tokens $(w_{a_1}, w_{a_2},\ldots,  w_{a_n})$ as parameters of the 
energy function, i.e., as coupling constants, which keep changing in each step that the LLM adds another word to 
the list to predict the next one, almost like an RG flow.

\section{Outlook: Scaling Laws and Beyond}

The basic ideas of deep learning (DL) are very old, going back to simple models for neurons in the 1940s \cite{McCulloch:1943cnb} 
and the more specific 
 DL techniques in the 1980s \cite{HintonBackProp}. Yet it was only in the 2010s that the full power became evident, thanks to the availability 
 of large enough data sets and increased computational power, see e.g.~\cite{Krizhevsky}. 
 The observation that DL works extremely well is largely empirical. 
 This raises the obvious question whether there is a theoretical framework that would explain this success, 
 somewhat analogous to how thermodynamics evolved in order to understand the behavior of  steam engines. 
The question is whether there are  really quantitatively precise observations in DL that are in need of a theory.  
For instance, LLMs produce interesting and surprisingly intricate text, but what should a theory of DL do about that? 
In the following I will briefly describe two kinds of quantitative empirical observations for which it might be beneficial to seek a theory: 
double descent in generalization curves  and scaling laws. 
(These were partly suggested by ChatGPT, cf.~the acknowledgements below.) 

The first  observation depends on the practice in DL to separate the data set into a \textit{training set} and a \textit{test set}. 
The training set is used to fix  the weights of the network as described above. The test set is then used 
to test the resulting network on new examples, i.e., to determine how well it generalizes beyond the training data. 
This in turn determines the \textit{test error}: 
the fraction of test examples for which the network does not give the desired result.  
 The \textit{generalization curve} now encodes how the test error varies as a function of the number of `parameters' (say the number 
 of neurons and hence the number of weights).  
One would expect a U-shaped curve: for few  parameters the network `underfits' and hence generalizes poorly to new 
 examples, leading to a large test error; increasing the number of parameters improves the performance until there are too many  
 parameters  and one overfits. Surprisingly, however, in modern DL one often dramatically overfits by having an enormous number of 
 parameters; in fact, the generalization curve is often not a  U-curve. Rather,  after the network describes  the 
 training set nearly perfectly, the test error at first gets worse but then improves again, leading to a `double descent' \cite{Belokin}. 
 One goal of a theory of DL should then be to predict the shape of the generalization curve or the location of the peak, 
 say as a function of the size of the dataset.

The second  empirical finding is  related  to large language models (LLMs)  and 
the observation of \textit{scaling laws}. We have encountered scaling laws in the discussion of phase transitions in sec.~\ref{scalinglaws}, where for instance the correlation length $\xi$ at a temperature close to the critical temperature $T^*$  
behaves as $\xi(T)\sim |T-T^*|^{-\nu}$ or the magnetization as $M(T)\sim |T-T^*|^{\beta}$, where $\nu, \beta, \ldots$ 
are collectively known as \textit{critical exponents}. Remarkably, the values of these critical exponents show a large 
degree of \textit{universality}: they are the same for many physical systems, independent of their detailed 
microscopic structure, which  thus becomes largely irrelevant close to a phase transition. 
Such scaling  laws are known as \textit{power laws} for the simple reason that the dependence of an observable $A$ on, say, 
the reduced temperature $t$, c.f.~(\ref{DEFt}), is of the form 
 \be\label{powerLAW} 
  A(t) = c\, t^{\alpha}\;, \qquad c, \alpha\in \mathbb{R}\;. 
 \ee
Equivalently, and often more conveniently,  the logarithms  $K:= \ln A$ and $\tau:=\ln t$ are related by an affine linear relation: 
 \be
  \ln A(t) = \alpha \,\ln t + \ln c \qquad \Leftrightarrow \qquad K(\tau) = \alpha\, \tau +c'\;. 
 \ee

  \begin{figure}
    \centering
    \includegraphics[width=0.99\textwidth]{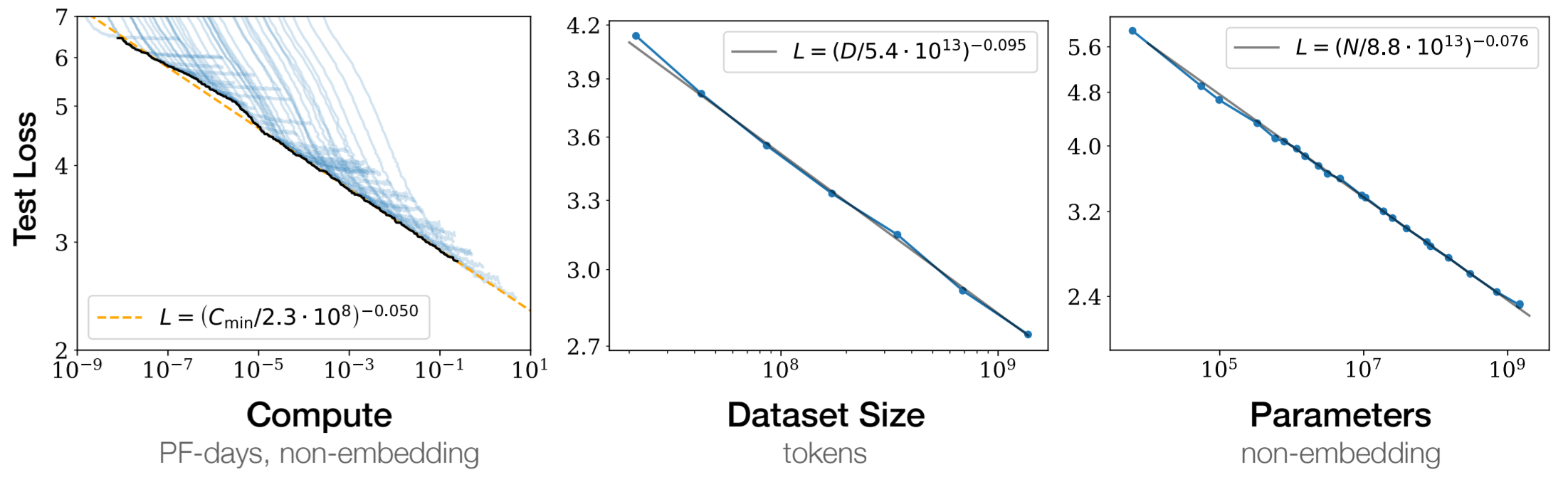}
    \caption{Scaling laws of large language models  (figure from   \cite{Kaplan:2020trn})  }
    \label{LLMScaling}
\end{figure}

It was found by Kaplan et.~al.~that the `performance'  of LLMs shows a similar power law dependence on the 
core parameters:  the number $N$ of trainable parameters, the size $D$ of the training data 
and the training compute $C$ \cite{Kaplan:2020trn}. More precisely, the so-called training loss $L$, which is the average of the 
loss or error function (as in (\ref{CostFunction})) over the training data, 
 has an approximate power law dependence (\ref{powerLAW}) on $N$, $D$ and $C$. For instance \cite{Kaplan:2020trn}, 
 \be
  L(N) \approx \left(\frac{N_C}{N}\right)^{\alpha_N} \;, \qquad \alpha_N\simeq 0.076\;, \quad N_C\simeq 8.8\cdot 10^{13}\;, 
 \ee  
with similar relations for $L(D)$ and $L(C)$, see figure \ref{LLMScaling}.  Remarkably, these relations show a degree of universality: 
they depend only weakly on the detailed deep learning architecture employed. 
They are thus similar to the critical phenomena in statistical physics, although the numerical universality is less striking. 

It is tempting to speculate about the possibility of a thermodynamics or statistical mechanics underlying neural networks 
and deep learning architectures that would allow one to better understand and predict such behaviors.

 \section*{Acknowledgments and Comments on these Notes }

These lecture notes are the result of an experiment: In early January 2026 I set myself the task to include in the remaining 
five to six weeks of my course on statistical physics, apart from the planned introduction to phase transitions and RG techniques, 
an introduction to neural networks and deep learning, even though I had only a basic    understanding  
of these topics. The goal was to use AI assistants such as 
 ChatGPT and Claude to understand the technical details (in addition, of course, to reading textbooks 
and original papers). I have frequently made use of AI assistants to understand technical constructions and to obtain 
missing steps in computations and proofs. Needless to say, every computation was carefully checked, and every sentence 
in this text was written by myself.

For assistance with teaching this course and preparing these notes 
I would like to thank Roberto Bonezzi,  Giuseppe Casale, 
Maria Kallimani and Camilla Lavino. 
Moreover, I would like to thank David Berman, Felipe Diaz  Jaramillo, 
Michael R.~Douglas and  Benjamin Lindner for discussions, comments 
on the manuscript, 
and for pointing me to some key references. 

\noindent

\appendix

\section{Basic Probability Theory}

\subsection{Probability Distributions}

We recall  basic concepts of classical probability theory, focusing on discrete probability distributions. (See, e.g., \cite{ProbBook} for a 
nice introduction.) 
We are given a finite set ${\cal X}$: the sample space or space of elementary events (or the configuration space in the main text).  
Its elements $x\in {\cal X}$  can be thought of as  possible outcomes of a game of chance or of an experiment. 
These elements are  `charged' with a real number between $0$ and $1$: its probability. This defines the probability distribution
 \be
  P : \,\mathcal{X} \ \rightarrow \ [0,1]\;, 
 \ee
which is normalized: 
 \be\label{normalizedPorb} 
  \sum_{x\in \mathcal{X}} P(x) = 1\;. 
 \ee
The simplest distribution is the \textit{uniform}  probability distribution, according to which 
all elementary events are equally likely, so that  (\ref{normalizedPorb}) 
implies  $P(x)=\frac{1}{n}$ for all $x\in\mathcal{X}$, where $n$ is the number of elements of ${\cal X}$. 
 
We now define an \textit{event} $A$ as a \textit{subset} of the sample  set: $A\subset {\cal X}$. For instance, for a fair die the sample  set 
is  $\mathcal{X}=\{1,2,3,4,5,6\}$, and the probability function is $P(1)=P(2)=\cdots =P(6)=\frac{1}{6}$. 
Special events include a particular number, e.g.~$A=\{5\}$, or the subset of odd numbers  $B=\{1,3,5\}$ (corresponding, respectively, 
to the throwing of a $5$ or the throwing of an odd number). Subsets $\{x\}$ with a single element are the 
\textit{elementary events}. 

One central goal of probability theory is to determine the probability $P(A)$ for any subset or  event $A\subset {\cal X}$. 
 This is simply the sum of the probabilities of its elements: 
  \be\label{probEvent} 
   P(A) = \sum_{x\in A} P(x)\;. 
  \ee 
 Note that the probability of the sample  set itself is $P(\mathcal{X})=1$ by (\ref{normalizedPorb}).  This expresses the certain fact 
 that the game of chance yields one of its possible outcomes (for it certainly cannot yield an impossible outcome).  
We can think, alternatively,  of $P$ as a probability measure  on ${\cal X}$ that assigns a number between $0$ and $1$
 to each subset.

\noindent 
\textit{Conditional Probabilities:}\\[0.5ex] 
The conditional probability is the probability of a certain event $A$ \textit{given} that another 
event $B$ has occurred.
To compute this we need to `transport' the probability distribution $P$ on $\mathcal{X}$ to a probability distribution $P_B$ 
on the subset $B\subset \mathcal{X}$. 
Thus, we need to give a rule that assigns to any subset $A\subset B$ a number $P_B(A)$ between $0$ and $1$. Since $A$ is then also 
a subset of $\mathcal{X}$ we have already such an assignment $P(A)$ from the original probability distribution, but we have to 
scale this number  in order to obtain  a properly normalized probability on $B$:\footnote{Note that for this 
definition we need to assume $P(B)>0$, i.e., that $B$ is a possible event. That better be the case, since we do not want to  condition 
the probability on an impossible event.}
 \be\label{ConditionalPROB} 
  P_B(A) := \frac{P(A)}{P(B)} \qquad \text{for} \qquad A\subseteq B \;. 
 \ee
Indeed, the normalization condition (\ref{normalizedPorb}) is then obeyed: 
 \be
  \sum_{x\in B} P_B(\{x\}) = \sum_{x\in B} \frac{P(\{x\})}{P(B)} = \frac{1}{P(B)}  \sum_{x\in B} P(x) = \frac{1}{P(B)} P(B)=1\;, 
 \ee
where we used (\ref{probEvent}).\footnote{The conditional probability distribution should not be confused with the marginal probability distribution  discussed around (\ref{Peff}) in the main text, even though in both one eliminates a certain subset of variables.  
For the conditional probability these variables are fixed and used to renormalize the distribution. 
For the marginal distribution these variables are summed over.}

In (\ref{ConditionalPROB}) we assumed $A\subseteq B$. It is convenient to generalize this formula to arbitrary sets $A\subseteq\mathcal{X}$. 
Of course, given the assumption 
that $B$ has occurred with certainty,  if $A$ has any elements that are not in $B$ their probability should be zero. 
This is implemented by replacing in (\ref{ConditionalPROB}) $A$ by its intersection with $B$:
  \be\label{TRUEConditionalPROB} 
  P(A|B) := \frac{P(A\cap B)}{P(B)}  \;. 
 \ee
This defines 
now a probability on the full space $\mathcal{X}$, albeit one that assigns zero probability to any elementary event that is not in $B$. 
Note that, in logical terms, $A\cap B$ denotes the event that $A$ AND $B$ has occurred. 
Correspondingly, as sets we have $A\cap B=B\cap A$, so that  $P(A\cap B)=P(B\cap A)$, which  implies 
with (\ref{TRUEConditionalPROB}) that 
 \be
P(A\cap B) = P(A|B) P(B) = P(B|A) P(A) 
 \;. 
 \ee
This  in turn implies  \textit{Bayes rule}: 
 \be
  P(A|B) = \frac{P(B|A) P(A)}{P(B)}  \;. 
 \ee

As an application we derive the `chain rule'  (\ref{chainrule}) used in the discussion of large language models in the main text. 
We consider the probability of a 
string of words or  tokens taken from a vocabulary  ${\cal V}$, focusing first 
on the set of strings consisting of two tokens. Thus, the sample space 
is ${\cal X}={\cal V}\times {\cal V}$, and we write a typical element as $(\sigma_1,\sigma_2)$, with $\sigma_1, \sigma_2\in {\cal V}$. 
For fixed tokens $x_1, x_2\in {\cal V}$ we then consider the events or subset 
 \be
 \begin{split}
  A_1 &:= \big\{ (x_1, \sigma_2)\big| \sigma_2\in {\cal V}\big\} \subset  {\cal X} \;, \\
  A_2 &:  = \big\{ (x_1, x_2)\big\}\subset  A_1  \subset  {\cal X}\;. 
 \end{split} 
 \ee 
The probability that the first token is $x_1$ is then $P(x_1):=P(A_1)$, the probability 
that the first two tokens are $(x_1,x_2)$ is  $P(x_1,x_2):=P(A_2)$, and the conditional probability 
that the second token is $x_2$ \textit{given} that the first token is $x_1$ is 
$P(x_2|x_1):=P(A_2|A_1)$. 
From  (\ref{TRUEConditionalPROB}) we infer 
 \be
  P(x_2|x_1)=P(A_2|A_1) 
  = \frac{P(A_2)}{P(A_1)}  = \frac{P(x_1,x_2)}{P(x_1)} \;, 
 \ee
and hence 
 \be
  P(x_1,x_2) = P(x_1) P(x_2|x_1) \;. 
 \ee
The general chain rule (\ref{chainrule})  follows inductively.

\subsection{Random Variables} 

A {random variable} or {observable}  is a real-valued function on the sample  set: 
 \be\label{randomvar} 
  {\cal O}:\, \mathcal{X} \ \rightarrow \ \mathbb{R}\;. 
 \ee 
(Of course we may also consider vector-valued and more complicated random variables.) The \textit{expectation value}  
of a random variable ${\cal O}$ is defined as 
 \be\label{expectatinvalue} 
  \langle {\cal O} \rangle = \sum_{x\in\mathcal{X}} {\cal O}(x) P(x)\;. 
 \ee
The expectation value can be viewed as a functional on random variables. This means that denoting the 
space of real-valued functions (\ref{randomvar}) by ${\cal F}_{\mathcal{X}}$, the expectation value is a map 
 \be
  \langle \;\cdot\;\rangle : \, {\cal F}_{\mathcal{X}} \ \rightarrow \ \mathbb{R}\;, 
 \ee 
that is linear and normalized: 
 \be\label{linearitynorm} 
  \langle a_1{\cal O}_1+ a_2{\cal O}_2\rangle  =  
  a_1\langle {\cal O}_1\rangle + a_2\langle {\cal O}_2\rangle \;, \qquad \langle 1\rangle  = 1\;. 
 \ee 
These properties  follow immediately from the definition (\ref{expectatinvalue}) and (\ref{normalizedPorb}).

The probability (\ref{probEvent}) of an event (i.e.~a subset) is the  expectation value of the characteristic function of the 
subset $A\subset\mathcal{X}$:
 \be
  \delta_{A}(x) : = \begin{cases}
      1  & \text{for  $x\in A$}\\
     0 & \text{else }
      \end{cases} \,, 
 \ee
for then we have with (\ref{expectatinvalue}) and (\ref{probEvent}) 
 \be
  \langle \delta_{A}\rangle = \sum_{x\in\mathcal{X}} \delta_{A}(x)  P(x)= \sum_{x\in A} P(x) = P(A) \;. 
 \ee

In the special case that the sample  set ${\cal X}$ is itself a set of real numbers we can define the random variable, often denoted 
by $X$, that yields back the corresponding number. For instance, for a fair die the sample  set  is $\mathcal{X}=\{1,2,3,4,5,6\}$, 
and we may define 
 \be\label{iderandomvare} 
  X:\mathcal{X}\rightarrow \mathbb{R}\;, \qquad X(i) = i\;, \qquad i=1,\ldots, 6\;. 
 \ee
The expectation value of the random variable $X$ then is equal to what is more conventionally called the arithmetic average: 
 \be
  \langle X\rangle = \sum_{x\in \mathcal{X}} X(x) P(x) = \sum_{i=1}^{6} X(i) P(i) = \frac{1}{6}  \sum_{i=1}^{6} i \;. 
 \ee 
Other statistical quantities can then be defined for $X$ or more general random variables ${\cal O}$; for instance, the 
\textit{variance} is defined by 
 \be
  {\rm Var}(\mathcal{O} ) := \langle ({\cal O} -\langle \mathcal{O}\rangle)^2\rangle = \langle {\cal O}^2\rangle 
  -\langle {\cal O}\rangle ^2\;, 
 \ee
where the second equality follows by writing out the square and using 
linearity and normalization (\ref{linearitynorm}) of the expectation 
value function  $\langle \,\cdot\,\rangle :  {\cal F}_{\mathcal{X}}  \rightarrow  \mathbb{R}$.  
The \textit{standard deviation} is defined as the square root of the variance.

\noindent 
\textit{Probability Distribution for Random Variable:}\\[0.5ex] 
In many statistical mechanics texts the notion of random variable is not explicitly used. 
The reason is that one may  define a new probability distribution on the space of \textit{values} of a random variable ${\cal O}$: 
 \be
  {\cal X}_{\cal O} : = \big\{y\,\big|\, \exists x\in {\cal X}:\;y={\cal O}(x)\big\}\;.   
 \ee 
The probability distribution $P$ on ${\cal X}$ may then be `transported' or `pushed-forward' to  a probability distribution 
on ${\cal X}_{\cal O}$:  
 \be\label{indicedprobdistr} 
   P_{\cal O} (y) \equiv \mathbb{P}({\cal O}=y) := P({\cal O}^{-1}(y)) = \sum_{ {\cal O}(x)=y}P(x)\;, 
 \ee
where $\mathbb{P}({\cal O}=y)$ is just a different yet more standard  notation, and  
  \be
   \qquad {\cal O}^{-1}(y):=\big\{x\in {\cal X}\,\big|\,  {\cal O}(x)=y \big\}\;. 
 \ee 
While we may thus restrict ourselves to talking about probability distributions, it often leads to more clarity 
by distinguishing between an underlying  probability space and the  random variable defined  on this space.

\subsection{Shannon Entropy}\label{Shannon} 

The entropy of a probability distribution $P$ on a finite set ${\cal X}$ is defined by 
 \be
  S[P] := -\sum_{x\in \mathcal{X}} P(x)\ln P(x) \;, 
 \ee
and has the following properties: 
 \begin{itemize}
  \item[i)] Since  $P(x)\in [0,1]$ we have $\ln P(x) \leq 0$ and hence 
   \be
    S[P]\geq 0 \;. 
   \ee 
 (For $P(x)=0$ we take $P(x)\ln P(x)=0$.) Equality $S[P]=0$ holds if and only if $P(x)=1$ for precisely one $x$ (and $P$ is zero 
 on all other elements). 
 \item[ii)] For the uniform probability distribution $\forall x\in \mathcal{X}: P(x)=\frac{1}{n}$, where $n$ is the number of 
 elements of $\mathcal{X}$, the entropy is 
  \be\label{uniformProp} 
   S = -\sum_{i=1}^{n} \frac{1}{n} \ln \frac{1}{ n} = \frac{1}{n} \ln n \sum_{i=1}^{n} 1 = \ln n\;. 
  \ee
 \item[iii)]  Denoting the uniform probability distribution under ii) by $P_U$ we have the inequality that 
 for \textit{any} probability distribution $P$ on the same set: 
  \be
   S[P ] \ \leq \ S[P_U]\;. 
  \ee
 The proof is a straightforward application of the inequality $\ln x\leq x-1$: 
  \be
  \begin{split} 
   S[P ] &= -\sum_{x} P(x)\ln P(x) = \sum_{x}  P(x) \ln \Big(\frac{n}{P(x)n} \Big) \\
   &=\ln n \sum_{x}P(x) + \sum_{x} P(x) \ln \Big(\frac{1}{P(x)n} \Big) \\
   &\leq \ln n  + \sum_{x} P(x)  \Big(\frac{1}{P(x)n}-1 \Big) \\
   &=  \ln n \\
   &= S[P_U]\;, 
  \end{split} 
  \ee
 where we used  (\ref{uniformProp}) in the last line.  
 \item[iv)] Given sample sets ${\cal X}_1$, with probability distribution $P_1$, and ${\cal X}_2$, with  probability distribution $P_2$, 
 the independent joint probability on ${\cal X}={\cal X}_1\times {\cal X}_2$, given by 
  \be
   P(x,y) = P_1(x) P_2(y)\;, 
  \ee
 has an entropy that is the sum of the entropies of $P_1$ and $P_2$: 
  \be
  \begin{split} 
   S[P_1P_2] &= -\sum_{(x,y)\in {\cal X}} P_1(x)P_2(y) \ln \big(P_1(x)P_2(y)\big) \\
   &= 
   -\sum_{(x,y)\in {\cal X}} P_1(x)P_2(y) \big( \ln P_1(x) +\ln P_2(y)  \big)  \\
   &= -\sum_{x\in {\cal X}_1} P_1(x)\ln P_1(x)\sum_{y\in \mathcal{X}_2} P_2(y)
   - \sum_{x\in {\cal X}_1} P_1(x)\sum_{y\in \mathcal{X}_2} P_2(y)\ln P_2(y) \\
   &= S[P_1] + S[P_2] \;. 
  \end{split} 
  \ee

 \end{itemize} 
 
Property  i) states that the entropy is always positive or zero and  is zero for maximal knowledge or minimal uncertainty. 
Properties ii) and iii) together imply that the entropy is maximal for minimal knowledge or maximal uncertainty (all events 
are equally likely). For a fair die the entropy is $S=\ln 6$ by (\ref{uniformProp}), and replacing it by a biased or rigged die 
\textit{decreases} the entropy (some events become  more likely, so the uncertainty is reduced).

There are two closely related notions of entropy that apply when two probability distributions, say  $P$ and $Q$, on the same 
 set ${\cal X}$ are compared. The \textit{cross entropy} is defined as 
 \be\label{DEFCross} 
  S(P,Q) :=  -\sum_{x\in {\cal X}} P(x)\ln Q(x) \;. 
 \ee 
Note that $S(P,P)=S(P)$. Furthermore, using again the inequality $\ln x\leq x-1$, we have 
 \be
 \begin{split} 
  S(P) - S(P,Q) &= \sum_x P(x) \ln \frac{Q(x)}{P(x)} 
    \leq \sum_x P(x)\Big(\frac{Q(x)}{P(x)}-1\Big) 
 = 0 \;, 
 \end{split} 
 \ee
and hence $S(P,Q)\geq S(P)$, with equality if and only if $P=Q$. 
The \textit{relative entropy}  or \textit{Kullback-Leibler divergence} is defined by 
 \be
  D(P|Q) := \sum_x P(x) \ln \frac{P(x)}{Q(x)}\;, 
 \ee
and hence obeys 
 \be\label{relativecross} 
  D(P|Q) = S(P,Q)-S(P)\;. 
 \ee 
The above inequality now implies $D(P|Q) \geq 0$ with equality if and only if $P=Q$. The relative entropy 
is thus a good measure for the `distance' between the two probability distributions $P$ and $Q$ (which, however, 
does not obey the axioms of a metric). In the main text this is applied for the case that $P$ is some given distribution, say obtained 
from a data set, and $Q$, depending on the parameters of a neural network, is a model of  this distribution. 
Learning the parameters by minimizing the relative entropy is equivalent to minimizing the cross entropy, since 
by (\ref{relativecross}) their difference is $S(P)$ and hence does not depend on the parameters.

\subsection{Example \& Continuous Probability Distribution}\label{RandomWalk}  
 We now consider a classic example of a probability distribution on a finite sample set and take the limit that the number of elements 
 is sent to infinity. This gives rise to a   \textit{continuous}  probability distribution.

 The example is that of throwing a fair  coin $N$ times, 
 each time giving head or tails, which for convenience we may represent as $-1$ and $+1$.  
 Thus, the sample set is  
  \be\label{samplecoins} 
   {\cal X}_N:=  \{-1,+1\}^N = \big\{ x=(x_1,\ldots, x_N)\,\big|\, x_i=\pm 1 \big\} \;. 
  \ee
 The number of elements or elementary events is $n=2^N$. By the assumption that the coin is fair, the probability distribution is 
 the uniform one: 
  \be
   P(x) = \frac{1}{2^N}\;.  
  \ee 
Since here we have defined the sample set as a set of real numbers we have the random variables defined as in (\ref{iderandomvare}), 
$X_i(x):=x_i\in\{- 1,+1\}$, and in terms of these we define 
 \be
  {\cal O}_N : = \sum_{i=1}^{N} X_i\;. 
 \ee
We can think of this random variable 
as encoding the outcome of a bet: If the coin gives tails we receive 1 Euro, if the coin gives head we have to pay 1 Euro, 
and ${\cal O}_N$  encodes our wins or losses after throwing the coin $N$ times. (Alternatively, we can think of this as a 
one-dimensional random walk, where at fixed time steps there is an equal chance that a particle moves to the 
right $(+1)$ or to the left $(-1)$. ${\cal O}_N$ then gives  the position of the particle after $N$ time steps.)  

The random variable ${\cal O}_N$ takes the possible values $m=k-(N-k)=2k-N$, where $k$ is the number of $x_i$ that 
equal $+1$ (so that the remaining $N-k$ variables equal $-1$). 
The probability distribution on this space of values of ${\cal O}$ is according to (\ref{indicedprobdistr}) given by 
 \be\label{mathcalPffd}
  \mathbb{P}({\cal O}=m) = \frac{1}{2^N} \binom{N }{k} 
  = \frac{1}{2^N}  \binom{N}{\frac{1}{2}(N+m) }
  \;, 
 \ee
where the binomial coefficient 
\be
 \binom{N }{k}   = \frac{N!}{(N-k)! k!} 
\ee
encodes the number of possibilities to choose $k$ variables  that are $+1$. 

One may take the large $N$ limit of (\ref{mathcalPffd}) by means of Stirling's formula, which states 
 \be\label{Stirling} 
  \ln N! \simeq N\ln N -N +\frac{1}{2} \ln (2\pi N) \quad \text{for large $N$} \;, 
 \ee
but the following trick of generating functions is a little simpler: For an auxiliary variable $t$ we define and compute 
 \be
 \begin{split} 
  G_N(t) &:= \big\langle e^{t{\cal O}_N}\big \rangle = \sum_{x\in {\cal X}_N}  e^{t{\cal O}_N(x) } P(x)\\
   &= \frac{1}{2^N}  \sum_{x\in {\cal X}_N}  e^{t\sum_ix_i}
   =  \frac{1}{2^N} \sum_{x_1=\pm 1} \cdots \sum_{x_N=\pm 1} e^{tx_1}\cdots e^{tx_N} \\
   &=\Big(\frac{1}{2} \sum_{x_1=\pm 1}e^{tx_1}\Big)^N
   = \left(\cosh t \right)^N\;. 
 \end{split}   
 \ee
Next, in order to take the large $N$ limit, we need to  rescale the random variable: 
  \be\label{ZNgenfunbct} 
  Z_N : =\frac{1}{\sqrt{N}} {\cal O}_N =  \frac{1}{\sqrt{N}} \sum_{i=1}^{N} X_i\;. 
 \ee
Its  corresponding generating function is then 
 \be
   G_N\big(\tfrac{t}{\sqrt{N}}\big) = \big\langle e^{tZ_N}\big \rangle =  \left(\cosh \big(\tfrac{t}{\sqrt{N}}\big)\right)^N\;. 
 \ee
We can now compute the large $N$ limit of the logarithm: 
 \be 
 \begin{split} 
  \lim_{N\rightarrow \infty} \ln   G_N\big(\tfrac{t}{\sqrt{N}}\big) 
  &= \lim_{N\rightarrow \infty} \, N\,  \ln \cosh \big(\tfrac{t}{\sqrt{N}}\big) \\
  &=  \lim_{N\rightarrow \infty} \, N\,  \Big( \tfrac{1}{2} \big(\tfrac{t}{\sqrt{N}}\big)^2 + {\cal O}\big(\tfrac{1}{N^2}\big) \Big) \\
  &= \tfrac{1}{2}t^2\;, 
 \end{split} 
 \ee
where we used the Taylor expansion $\ln \cosh u = \tfrac{1}{2} u^2+\cdots$.  
Thus, taking the exponential we have shown 
 \be
  G_N\big(\tfrac{t}{\sqrt{N}}\big) = \big\langle e^{tZ_N}\big \rangle   \ \simeq  \ e^{ \frac{1}{2}t^2} \quad\text{for large $N$} \;. 
 \ee
We now ask: What is the \textit{continuous} probability distribution $\rho(z)$ for a  continuous random variable $z$ 
that is compatible with the above large $N$ limit in  that 
 \be\label{dfvdfg} 
   \big\langle e^{t z}\big \rangle  = \int_{-\infty}^{\infty} dz \,e^{tz}\, \rho(z) = e^{ \frac{1}{2}t^2}\;\; ? 
 \ee 
The answer is the Gaussian distribution 
 \be\label{Gaussian} 
  \rho(z) = \frac{1}{\sqrt{2\pi}} \,e^{-\frac{z^2}{2}} \;.  
 \ee 
This follows quickly by inserting (\ref{Gaussian}) into (\ref{dfvdfg}), completing the square in the exponential and 
using the standard Gaussian integral $\int_{-\infty}^{\infty} dz \,e^{-\frac{z^2}{2}}  = \sqrt{2\pi}$.

\end{document}